\documentclass{aa}  

\usepackage{graphicx}
\usepackage{txfonts}
\usepackage{lipsum}
\usepackage{subcaption}         % necessary for continued figures, example in section 3
\usepackage{lscape}             % to rotate a single page table, example in appendix.
\usepackage{placeins}           % useful with \FloatBarrier, to keep 
\begin{document}

   \title{Interpretation of fast radio bursts from CHIME/FRB baseband data}

   \subtitle{}

%%%%%%%%%%%%%%%%%%%%%%%%%%%%%%%%%%%%%%%%
% Please do not include ORCIDs next to author names.
% Only ORCIDs authenticated by individual authors in EDP Sciences editorial system will be taken into account.
% ORCIDs included here will be removed.
%%%%%%%%%%%%%%%%%%%%%%%%%%%%%%%%%%%%%%%%

   \author{Wen-Qi Ma\inst{1,2}
        \and Zhi-Fu Gao\inst{1}\thanks{Corresponding author: zhifugao@xao.ac.cn}
        \and Xiang-Han Cui\inst{2,3}
        \and Ju-Mei Yao\inst{1}
        \and Biao-Peng Li\inst{4}
        \and Fa-Yin Wang\inst{5,6}\fnmsep
        }

   \institute{Xinjiang Astronomical Observatory, Chinese Academy of Sciences, Urumqi, Xinjiang 830011, China\\
             % \email{}
             %\thanks{Shows the usage of elements in the author field}
            \and School of Astronomy and Space Science, University of Chinese Academy of Sciences, Beijing 100049, China\\
            \and National Astronomical Observatories, Chinese Academy of Sciences, Beijing 100012, China\\
            \and College of Science, Hunan Institute of Science and Engineering, Yongzhou, Hunan 425199, China\\
            \and School of Astronomy and Space Science, Nanjing University, Nanjing 210023, China\\
            \and Key Laboratory of Modern Astronomy and Astrophysics (Nanjing University), Ministry of Education, Nanjing 210093, China}

   \date{Received September 10, 20xx}

% \abstract{}{}{}{}{}
% 5 {} token are mandatory
 
  \abstract
%   % context heading (optional)
%   % {} leave it empty if necessary  
%    {Optional, leave empty if necessary.  The heading “Context” is used when needed to
% give background information on the research conducted in the paper}
%   % aims heading (mandatory)
%    {Mandatory. The objectives of the paper are defined here.} 
%   % methods heading (mandatory)
%    {Mandatory. The methods of the investigation are outlined here}
%   % results heading (mandatory)
%    {Mandatory. The results are summarised here.}
%   % conclusions heading (optional), leave it empty if necessary
%    {Optional, leave empty if necessary.  “Conclusions” can be used to
% explicit the general conclusions that can be drawn from the paper.}
    {Fast radio bursts (FRBs) are radio pulses that originate from cosmological distances. Thousands of FRB sources have been discovered, but their origins remain unknown. Analyses of the energy function and redshift evolution of the volumetric event rate could provide crucial insights into FRB progenitors. In this paper, we used the $V_\mathrm{max}$ method to analyse the FRB energy functions and volumetric event rate evolution based on CHIME/FRB BaseCat 1 data. The baseband data provide improved fluence and bandwidth measurements compared to Catalogue 1, enabling a more robust analysis of FRB energy and population studies. The $V_\mathrm{max}$ method allows us to investigate volumetric event rate evolution as a function of redshift without any prior assumptions. The selection functions for baseband data were derived to re-weight the samples when calculating the number density. We observe Schechter-like shapes for the energy function with a unified slope index of $\gamma \in [-1.8, -1.6]$. Neither the cosmic star formation rate density (SFRD) nor the cosmic stellar-mass density scenario can be ruled out. We further discuss the scenario that FRBs are primarily produced from systems that evolve with a delay relative to SFRD. The rate evolution trend suggests a scenario of a delayed model with a delay time $\tau \approx 4.66$ Gyr, hinting at delayed core-collapse supernovae or compact binary mergers. Our work highlights the importance of baseband data, and the BaseCat 2 data are required to explore FRB populations and rate evolution.}
    
   \maketitle

\nolinenumbers

%%%%%%%%%%%%%%%%%%%%%%%%%%%%%%%%%%%%%%%%%%%%%%%%%%%%%%%%%%%%%%
\section{Introduction}

Fast radio bursts (FRBs) are bright signals from extragalactic distances, characterised by a short duration, high dispersion measure (DM), and high brightness temperature ($T_b$). The first FRB signal was identified in archival data from the Parkes telescope \citep{2007Sci...318..777L}. The first repeater source, FRB~20121102A, was discovered by the Arecibo Telescope \citep{2014ApJ...790..101S, 2016Natur.531..202S}. Follow-up observations localised it to a dwarf galaxy at $z = 0.19$, confirming its cosmological origin \citep{2017ApJ...834L...7T}.

Fast radio burst sources and their bursts have been detected across a wide frequency range (from hundreds of megahertz to several gigahertz) \citep{2018ApJ...863....2G, 2021Natur.596..505P}. More than 3000 sources have been discovered \citep{2026ApJS..283...34C}. Fast radio bursts are typically classified as repeaters and non-repeaters. Only a few sources have been observed to repeat \citep{2021ApJ...906L...5A}. However, no consensus exists on populations and progenitors, and it has not been determined whether repeaters and non-repeaters have similar origins \citep{2021ApJ...911L...3P, 2021MNRAS.500.3275C}. A recent analysis of FRB host galaxies has revealed that FRBs originate from diverse galaxies and local environments, suggesting complex origins \citep{2020ApJ...895L..37B, 2022AJ....163...69B}. Numerous models have been proposed to explain the production of repeaters and non-repeaters \citep{2023RvMP...95c5005Z}. It is confirmed that the repeater source, FRB~20200428, was associated with X-ray bursts from the galactic magnetar, SGR~1935+2154, suggesting a possible origin from active magnetars \citep{2020Natur.587...59B, 2020Natur.587...54C, 2020ApJ...898L..29M, 2021NatAs...5..378L, 2021NatAs...5..372R, 2021NatAs...5..401T}. The estimated isotropic-equivalent energy of FRB~20200428 ($E_\mathrm{iso}\sim 10^{34}--10^{35}~\mathrm{erg}$) is several thousand times greater than that of any radio pulse from the brightest galactic radio source, the Crab pulsar \citep{2020Natur.587...59B, 2020Natur.587...54C}, exhibiting the energetic nature of FRBs. It is noted that FRB~20200428 is fainter than other FRBs by several orders of magnitude. Whether FRBs from cosmological distances could also originate from magnetars remains unclear.

The Canadian Hydrogen Intensity Mapping Experiment (CHIME) radio telescope has discovered more than 3000 FRBs with its excellent detection capability \citep{2019Natur.566..235C, 2019ApJ...885L..24C, 2019Natur.566..230C, 2021ApJS..257...59C, 2023ApJ...947...83C}. The first CHIME/FRB catalogue (hereafter Catalogue 1) was published in 2021, including 536 bursts (474 newly discovered non-repeater bursts and 62 bursts from 18 previously reported repeaters) observed between 400 and 800 MHz from 2018 July 25 to 2019 July 1 \citep{2021ApJS..257...59C}. Investigation of a large homogeneous FRB sample helps mitigate the systematic uncertainty from different telescopes. The CHIME/FRB Catalogue 1 is an efficient dataset to investigate FRB properties and provides valuable opportunities for investigating FRB population and origins.

Investigating FRB populations will provide insights into their origins. A fundamental way to investigate FRB populations is to study their luminosity and energy functions. The volumetric rate can be derived from luminosity or energy functions and compared with other astrophysical events to examine FRB origins \citep{2019NatAs...3..928R, 2020MNRAS.494.2886H, 2020MNRAS.494..665L, 2022MNRAS.509.4775J, 2022ApJ...924L..14Z, 2022MNRAS.511.1961H, 2022Ap&SS.367...66C, 2023ApJ...944..105S, 2025PASA...42....3A}. The research methods for luminosity or energy functions are equivalent in many circumstances. \citet{1997AJ....114..898W} summarised the methods for calculating the luminosity function of galaxies. These include the Lynden-Bell's $c^{-}$ method \citep{1971MNRAS.155...95L}, the $V_\text{max}$ method \citep{1968ApJ...151..393S}, and the Bayesian method \citep{2013MNRAS.431..874C}. Non-parametric methods (i.e. the $V_\text{max}$ method and the Lynden-Bell's $c^{-}$ method) require certain uniformity in the data. By contrast, Bayesian estimation methods are more flexible in modelling systematic errors and thus have fewer restrictions based on sample conditions. Using the $V_\text{max}$ method, \citet{2022MNRAS.511.1961H} analysed energy functions of FRBs from CHIME/FRB Catalogue 1 data and report energy function slope indices $\alpha = -1.4^{+0.7}_{-0.5}$ and $\alpha = -1.1^{+0.6}_{-0.4}$ for different redshift bin settings. \citet{2023ApJ...944..105S} utilised the CHIME/FRB Catalogue 1 data to derive the intrinsic energy distribution of FRBs by calibrating observational selection effects \citep{2021ApJS..257...59C} and modelling through Bayesian statistics based on the method proposed by \citet{2022MNRAS.509.4775J}.

The relations between FRB volumetric event rate evolution and cosmic stellar evolution could provide information about FRB populations. Analyses of FRB populations show diverse results. \citet{2020MNRAS.498.3927H} studied the time-integrated energy functions and volumetric rates for both repeaters and apparently non-repeaters and find no significant redshift evolution for apparent non-repeaters, while the rate for repeaters may increase with redshift. \citet{2022MNRAS.510L..18J} concluded that the FRB rate evolves with redshift consistently with, or faster than, the star formation rate using large samples of FRBs detected by ASKAP and Parkes. \citet{2022MNRAS.511.1961H} used 164 apparently non-repeater samples from the first CHIME/FRB Catalogue to investigate redshift evolution. They argued that the apparent non-repeaters do not evolve with the cosmic star formation rate and that the evolution is likely controlled by old populations such as neutron stars and black holes. \citet{2022ApJ...924L..14Z} rule out the hypothesis that FRBs track the cosmic star-formation history and discuss the possibility of either a delayed or hybrid model for FRB source models. \citet{2023ApJ...944..105S} also discuss the contribution of host galaxy dispersion and the implications of cosmological evolution and conclude that the CHIME/FRB source rate evolves with the cosmic star formation rate. This is consistent with the results from \citet{2022MNRAS.509.4775J}. \citet{2024ApJ...973L..54C} find that the rate of non-repeating FRBs declines monotonically with redshift, similar to that of short gamma-ray bursts (sGRBs), using the Lynden-Bell's $c^{-}$ method. They suggest that neutron stars and black holes are closely linked to the origins of FRBs.

However, the flux density and fluence values of most non-repeaters to date represent lower limits, with the exception of ASKAP FRBs \citep{2018Natur.562..386S, 2017ApJ...841L..12B}. For example, flux density and fluence values from Catalogue 1 not only have large uncertainties but are also interpreted as lower limits due to the limited positional accuracy of FRBs in Catalogue 1 \citep{2021ApJS..257...59C}. Therefore, the CHIME/FRB Collaboration released the first CHIME/FRB baseband data (hereafter BaseCat 1) \citep{2024ApJ...969..145C}. The baseband system improves time resolution, burst localisation, as well as flux density and fluence measurements. No correlations were found between scattering time, DM, rotation measure (RM), and the linear polarisation fraction \citep{2025ApJ...979..160S}. Baseband data highlight the importance of high time resolution for understanding FRB morphology, emission mechanisms, and origins. In addition, more precise data allow us to explore FRB cosmology in greater depth. By comparing the CHIME/FRB baseband data with the optical galaxy catalogue, two potential host galaxies and a marginal candidate host galaxy were confirmed, helping identify the morphologies of the FRB host galaxy \citep{2024ApJ...961...99I}.

In this paper, we discuss the energy function and volumetric rate evolution as a function of redshift for FRB populations derived from updated CHIME/FRB baseband data. The aims of our research are: (1) discussing the differences between baseband data and Catalogue 1 data, as well as the effects on the analysis, (2) calculating the energy function of FRBs using updated values, and (3) discussing FRB populations and progenitor models by examining the volumetric event rate evolution. The paper is organised as follows. In Sect.~\ref{sec:data}, we introduce the baseband data and data selection process, as well as the redshift and energy estimation. In Sect.~\ref{sec:method}, the selection functions for several properties and the volumetric rate estimation method are introduced. In Sect.~\ref{sec:results}, we present our results of the energy functions and volumetric rates within different redshift bins. The limitations of the data and the effects of measurement uncertainties are discussed in Sect.~\ref{sec:discussion}. Our conclusions are summarised in Sect.~\ref{sec:conclusion}.

%%%%%%%%%%%%%%%%%%%%%%%%%%%%%%%%%%%%%%%%%%%%%%%%%%%%%%%%%%%%%%
\section{Data}
\label{sec:data}

Baseband data (i.e. voltage data) measured by radio telescope receivers contain information about both the intensity and phase of signals. Large field-of-view surveys typically reduce the data rate before searching for FRBs due to the large amount of baseband data and the cost of processing. Similar to many new-generation telescopes, CHIME is able to store baseband data for each frequency channel when an FRB candidate is identified. The offline pipelines are subsequently used to process the baseband data at a later stage \citep{2018ApJ...863...48C}. The baseband data can improve the beam lobe pointing towards the source position. Due to the improved beam model, the baseband data demonstrate superior accuracy of localisation and other key parameters.

\begin{figure*}[!ht]
    \sidecaption
    \includegraphics[width=12cm]{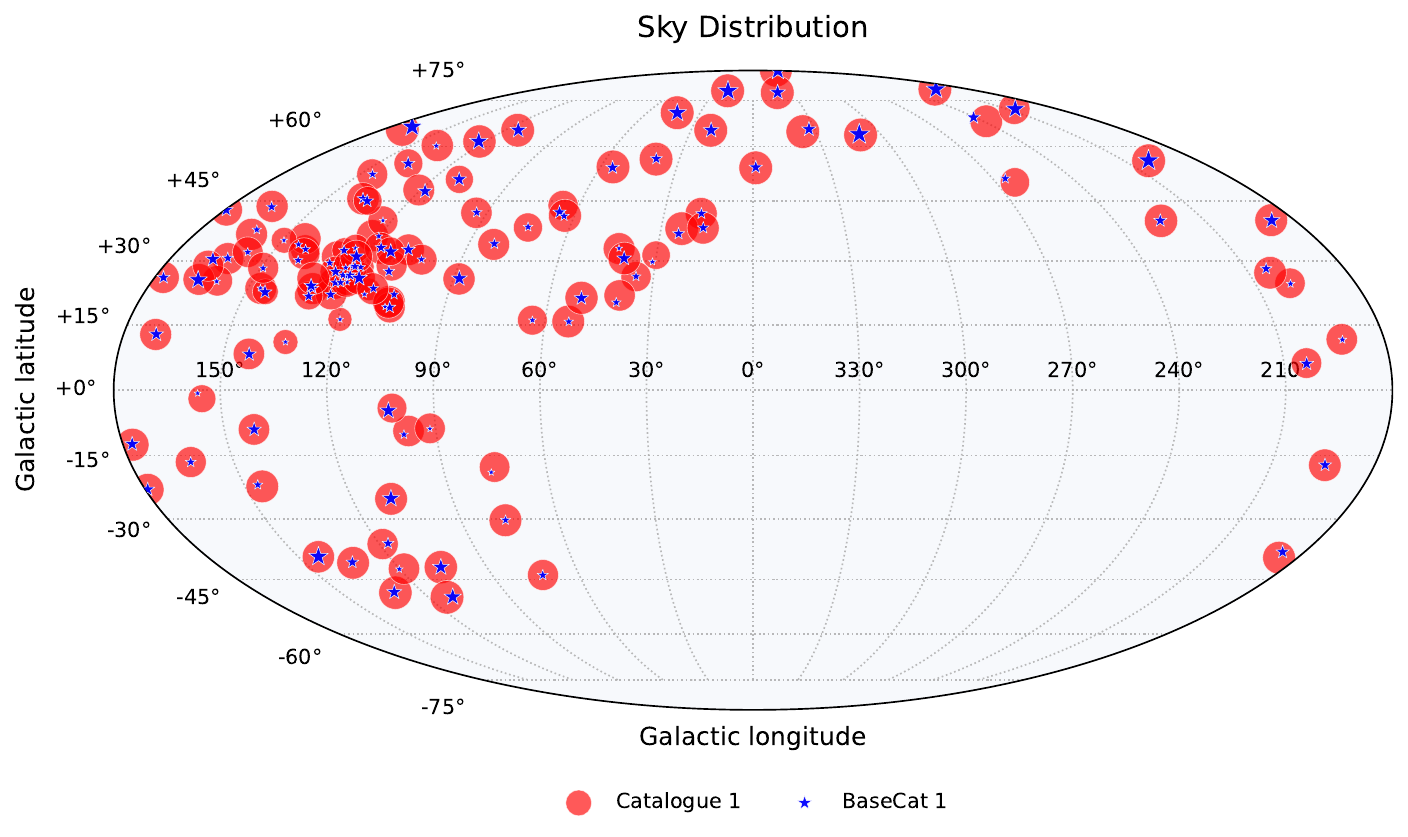}
    \caption{Sky distribution of CHIME/FRB Catalogue 1 and BaseCat 1 sources. The galactic coordinates of BaseCat 1 sources were calculated from their right ascensions and declinations. Note that the marker sizes are not the true values, and are rescaled to visually show the importance of localisation accuracy.}
    \label{fig:coordinates}
\end{figure*}

Unfortunately, only sources observed between 2018 December 9 and 2019 July 1 during the Catalogue 1 observation epoch have available baseband data \citep{2024ApJ...969..145C}. CHIME/FRB BaseCat 1 contains 140 bursts, including 128 apparently non-repeater bursts and 12 repeater bursts. The preliminary data selection criteria discard sources with no available flux density, fluence or bandwidth values. Typically, only the first-detected repeat burst from each repeater is kept in the analysis. There are 11 out of 12 repeater bursts from already recognised repeaters (FRB~20180916B, FRB~20190116B, FRB~20190222A, FRB~20190604A, FRB~20180814A and FRB~20180908B). FRB~20190117A is the first detected burst of the source. FRB~20190116A and FRB~20190116B are repeater bursts from the same source. However, CHIME has additional detection sensitivity to repeaters; thus, the assumption of identical detection biases for repeaters and non-repeaters is unlikely to be true \citep{2021ApJS..257...59C, 2023ApJ...944..105S}. Considering these effects, repeater bursts were discarded. In addition, FRB~20190613B was removed due to its low DM (less than $1.5\times\mathrm{DM_{MW}}$). Finally, 122 bursts remained for analysis after applying the data selection criteria.

\begin{figure}[!ht]
\centering
\includegraphics[width=\hsize]{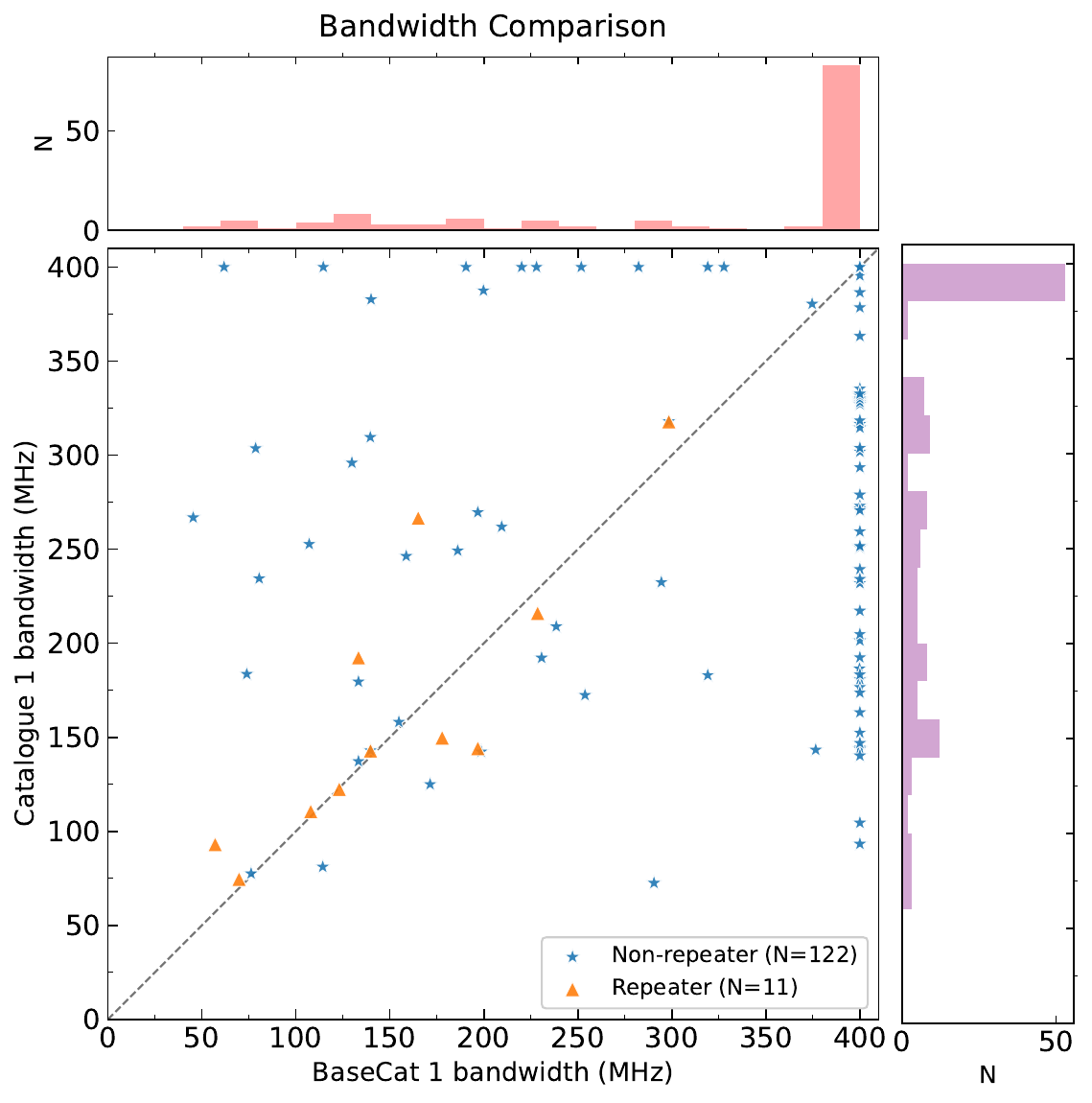}
  \caption{Bandwidth comparison between BaseCat 1 and Catalogue 1 data. The blue stars and yellow triangles represent non-repeaters and repeaters. Top and right subplots: Histograms of the bandwidths of BaseCat 1 and Catalogue 1 sources. For sources with no specific bandwidth, the lower limit values are adopted.}
     \label{fig:bandwidth}
\end{figure}

Figure~\ref{fig:coordinates} demonstrates the galactic coordinates and mean errors of sources from BaseCat 1 and Catalogue 1. The errors of galactic longitude (G.L.) and galactic latitude (G.B.) are transformed from celestial coordinates (R.A. and Dec.), and the mean errors are calculated from the root mean square of G.L. and G.B. errors. The marker sizes represent the coordinate errors. Note that the marker sizes are not the true values, and are rescaled to visually show the importance of localisation accuracy. Accurate sky positions enable precise beam forming, facilitating robust measurements of bandwidths. Figure~\ref{fig:bandwidth} illustrates the bandwidth comparison between sources from Catalogue 1 and BaseCat 1. BaseCat 1 sources appear to be more broadband than Catalogue 1 sources. Nearly 80\% of the bursts are emitted across the whole frequency band. The differences are primarily due to an uncorrected instrumental bandpass response, which emphasises the importance of baseband data for spectral property analysis \citep{2021ApJS..257...59C}. In contrast, repeater bursts show no significant change in bandwidth, remaining narrow-band emission.

\begin{figure*}
    \centering
    \includegraphics[width=\hsize]{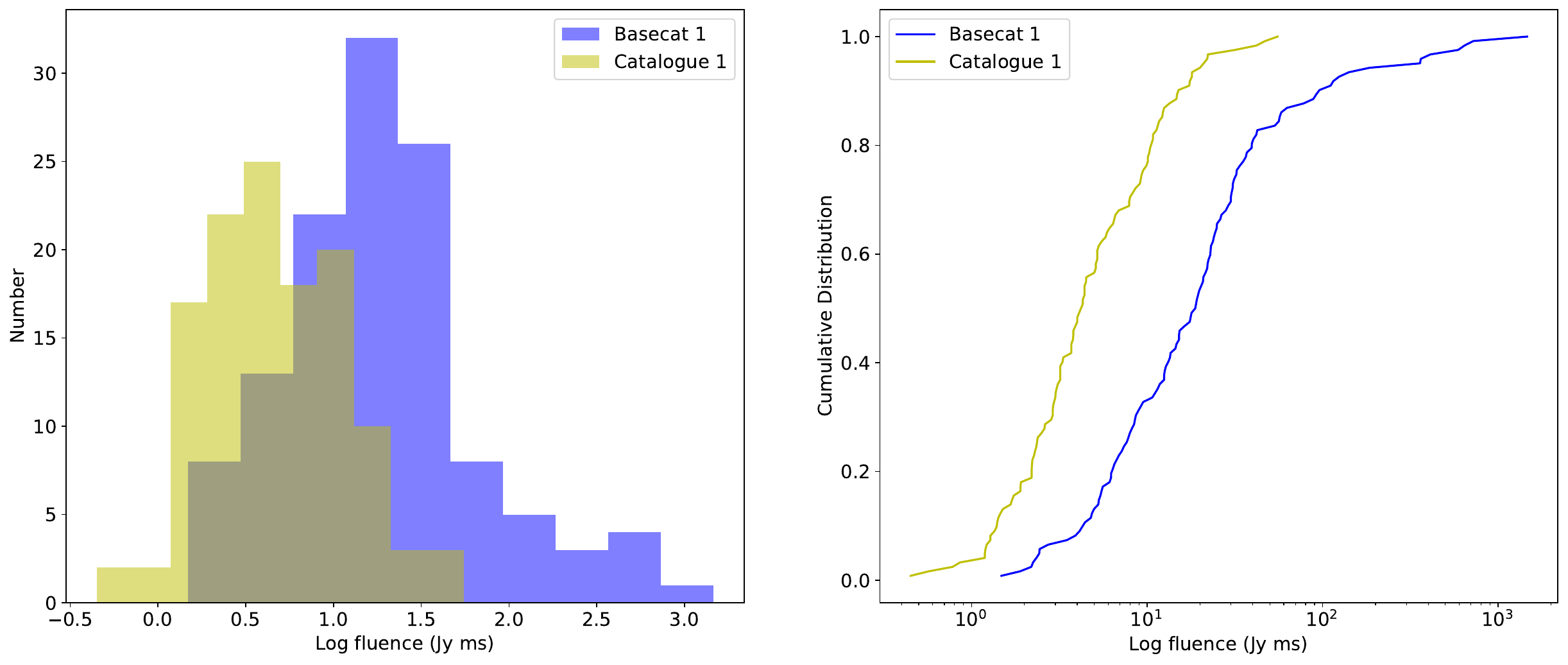}
    \caption{Fluence value comparison (left panel) and the normalised cumulative distribution (right panel) between BaseCat 1 (blue) and Catalogue 1 (yellow) data.}
    \label{fig:fluence}
\end{figure*}

The improved localisations help correct the beam response of the instrument and measure flux density and fluence with higher accuracy \citep{2024ApJ...969..145C}. A statistical analysis reveals systematic differences in flux density and fluence values between BaseCat 1 and Catalogue 1 sources. This discrepancy is visually demonstrated in Figure~\ref{fig:fluence}. The median values of fluence from Catalogue 1 and BaseCat 1 are 4.25 Jy ms and 18.9 Jy ms, respectively. The maximum fluence value from BaseCat 1 exceeds 1000 Jy ms, which is ten times larger than the value from Catalogue 1. The larger fluence measurements and wider bandwidths lead to a lack of low-fluence and low-energy events. Considering the fluence completeness and the larger uncertainty of low-fluence events, the threshold of 5 Jy ms was adopted. Events below the fluence threshold were discarded, leaving 106 events for further analysis.

\subsection{Redshift estimation}
\label{subsec:redshift}

Since FRBs originate from cosmological galaxies, it is hard to identify their host galaxies. Thus, we estimated the pseudo redshift from observed DM values using Bayesian estimation. The higher time resolution of the baseband data significantly improves the DM values \citep{2024ApJ...969..145C}. The FRB DM contributions consist of many components: the interstellar medium (ISM) in the Milky Way ($\mathrm{DM_{MW}}$), the outer matter halo of the Milky Way ($\mathrm{DM_{halo}}$), the intergalactic medium ($\mathrm{DM_{IGM}}$), the host galaxy ($\mathrm{DM_{host}}$), and the source ($\mathrm{DM_{source}}$). The DM contribution without including the Milky Way component ($\mathrm{DM_{ex}}$) can be described as \citep{2021ApJ...906...49Z, 2023RvMP...95c5005Z}

\begin{align}
    \mathrm{DM_{ex}} &= \mathrm{DM_{FRB}}-\mathrm{DM_{MW}}-\mathrm{DM_{halo}} \notag \\
    &= \mathrm{DM_{IGM}}+\frac{\mathrm{DM_{host}}+\mathrm{DM_{source}}}{1+z},
\label{eq:01}
\end{align}

\noindent where the values of $\mathrm{DM_{MW}}$ and $\mathrm{DM_{halo}}$ are calculated from the YMW16 \citep{2017ApJ...835...29Y} and YT20 \citep{2020ApJ...888..105Y} models. The denominator $(1+z)$ reflects the frequency stretching effect caused by the dispersion of the medium in the host galaxy due to cosmological expansion (i.e. $\mathrm{DM_{host}}$ and $\mathrm{DM_{source}}$ must be corrected from the observer frame to the host-galaxy frame).

The redshift of an FRB is primarily determined by $\mathrm{DM_{IGM}}$, whose measurement is challenging. The distribution of the intergalactic medium at the specific redshift can be described as \citep{2020Natur.581..391M, 2021ApJ...906...49Z}

\begin{equation}
    P\left (\frac{\mathrm{DM_{IGM}}}{\langle\mathrm{DM_{IGM}}\rangle}\right ) = A\times\left(\frac{\mathrm{DM_{IGM}}}
{\langle\mathrm{DM_{IGM}}\rangle}\right)^{-\beta}e^{-\frac{\left[\left(\frac{\mathrm{DM_{IGM}}}{\langle\mathrm{DM_{IGM}}\rangle}\right)^
{-\alpha}-C_0\right ]^2}{2\alpha^2\sigma_\mathrm{DM}^2}}, \quad\Delta>0,
\label{eq:02}
\end{equation}

\noindent where $A$ is the normalisation parameter and $\alpha=\beta=3$ corresponds to the slope and the density profile of the gas in halos. Here, $\sigma_\mathrm{DM}$ represents the standard deviation of DM, and $C_0$ is a free parameter. The parameter values were adopted from \citet{2021ApJ...906...49Z}. The averaged DM term ($\langle\mathrm{DM_{IGM}}\rangle$) relates to the cosmological redshift, which is given by \citep{2020Natur.581..391M}:

\begin{equation}
    \langle\mathrm{DM_{IGM}}\rangle=\frac{3c\Omega_bH_0}{8\pi G m_p}\int_{0}^{z}\frac{(1+z^\prime)f_\mathrm{IGM}(z^\prime)
f_e(z^\prime)}{\sqrt{{(1+z^\prime)}^3\Omega_m+\Omega_\Lambda}}\mathrm{d}z^\prime,
\label{eq:03}
\end{equation}

\noindent where $H_0$ is the Hubble constant, $m_p$ is the proton mass, and $f_\mathrm{IGM}$ represents the fraction of baryon mass in the intergalactic medium \citep{2012ApJ...759...23S}. The electron fraction is given by $f_e=Y_\mathrm{H}X_{e,\mathrm{H}}(z)+\frac{1}{2}Y_\mathrm{He}X_{e,\mathrm{He}}(z)$, where $Y_\mathrm{H} = 3/4$ and $Y_\mathrm{He} = 1/4$ are the mass fractions of hydrogen and helium, respectively. The values $X_{e,\mathrm{H}}(z)$ and $X_{e,\mathrm{He}}(z)$ denote the ionisation fractions of intergalactic hydrogen and helium. The cosmological parameters were adopted from Planck 2018 results \citep{2020A&A...641A...6P}: $H_0=67.74$~$\mathrm{km~s^{-1}~{Mpc}^{-1}}$, $\Omega_m=0.3111$, and $\Omega_\Lambda=0.6889$.

The contributions from the FRB host galaxies and source local environments remain poorly constrained owing to the complicated host-galaxy morphology and inclination, as well as to uncertain origins. For a rough analysis, we combined these two components into a single $\mathrm{DM_{host}}$ term. We modelled the $\mathrm{DM_{host}}$ distribution as a lognormal distribution, following recent studies \citep{2020ApJ...900..170Z, 2023MNRAS.518..539M, 2025ApJ...981...24M}:

\begin{equation}
    P(\mathrm{DM_{host}})=\frac{1}{\sqrt{2\pi}\sigma_\mathrm{host}\mathrm{DM_{host}}}e^{-\frac{(\mathrm{ln~DM_{host}-\mu_{host})}^2}{2\sigma_\mathrm{host}^2}}.
\label{eq:04}
\end{equation}

\noindent \citet{2020ApJ...900..170Z} provide the following redshift-evolution result: $\mu = \mathrm{ln}~(32.97{(1 + z)}^{0.84})$ and $\sigma_{\mathrm{host}} = 1.248$ for non-repeaters from the IllustrisTNG simulation. \citet{2023MNRAS.518..539M} studied the $\mathrm{DM_{host}}$ contribution from the assumption that FRBs track the star formation rate or stellar mass of host galaxies. They conclude that $\mathrm{DM_{host}}$ may increase as host galaxy mass increases. They also argue that $\mathrm{DM_{host}}$ could evolve with redshift. \citet{2025ApJ...981...24M} identify no significant difference in $\mathrm{DM_{host}}$ between repeaters and non-repeaters from 49 localised FRBs. The best-estimated parameters are $\mu_{\mathrm{host}} = 66.63^{+17.67}_{-14.81}$ and $\sigma_{\mathrm{host}} = 1.57^{+0.21}_{-0.18}$ for non-repeaters, and $\mu_{\mathrm{host}} = 64.46^{+18.05}_{-15.39}$ and $\sigma_{\mathrm{host}} = 1.09^{+0.26}_{-0.18}$ for repeaters. Therefore, the values of $\mu_{\mathrm{host}} = 65.0$ and $\sigma_{\mathrm{host}} = 1.00$ are assumed for theoretical analysis. 

We estimated the redshift posterior distribution for an FRB with a measured $\mathrm{DM_{ex}}$ based on the Bayesian theorem:

\begin{equation}
    P(z|\mathrm{DM_{ex}})\propto P(z)L(\mathrm{DM_{ex}}|z),
\label{eq:05}
\end{equation}

\noindent where the likelihood function $L(\mathrm{DM_{ex}}|z)$ is defined for a given $z$. Since $\mathrm{DM_{IGM}}$ and $\mathrm{DM_{host}}$ are observationally coupled, the likelihood function can be written as

\begin{align}
    L(\mathrm{DM_{ex}}|z) &= \int_0^{\mathrm{DM_{ex}}}P(\mathrm{DM_{IGM}}|z)P(\mathrm{DM_{host}}|z) \mathrm{d~DM_{host}} \notag \\
    &= \int_0^{\mathrm{DM_{ex}}}P(\mathrm{DM_{ex} - DM_{host}}|z)P(\mathrm{DM_{host}}|z)\mathrm{d~DM_{host}}.
\label{eq:06}
\end{align}

We used the Python package \texttt{emcee}\footnote{\url{https://github.com/dfm/emcee}} for Markov chain Monte Carlo (MCMC) sampling \citep{2013PASP..125..306F}, and the prior distribution was chosen as a uniform distribution of $z\in[0, 3]$ for each source. The best-estimated redshift and uncertainty were adopted from the median values and $\pm$1$\sigma$ posterior ranges after running 20 walkers for 2000 steps (discarding the first 200 steps as burn-in).

\subsection{Energy calculation}
\label{subsec:energy}

\begin{figure}[!ht]
\centering
\includegraphics[width=\hsize]{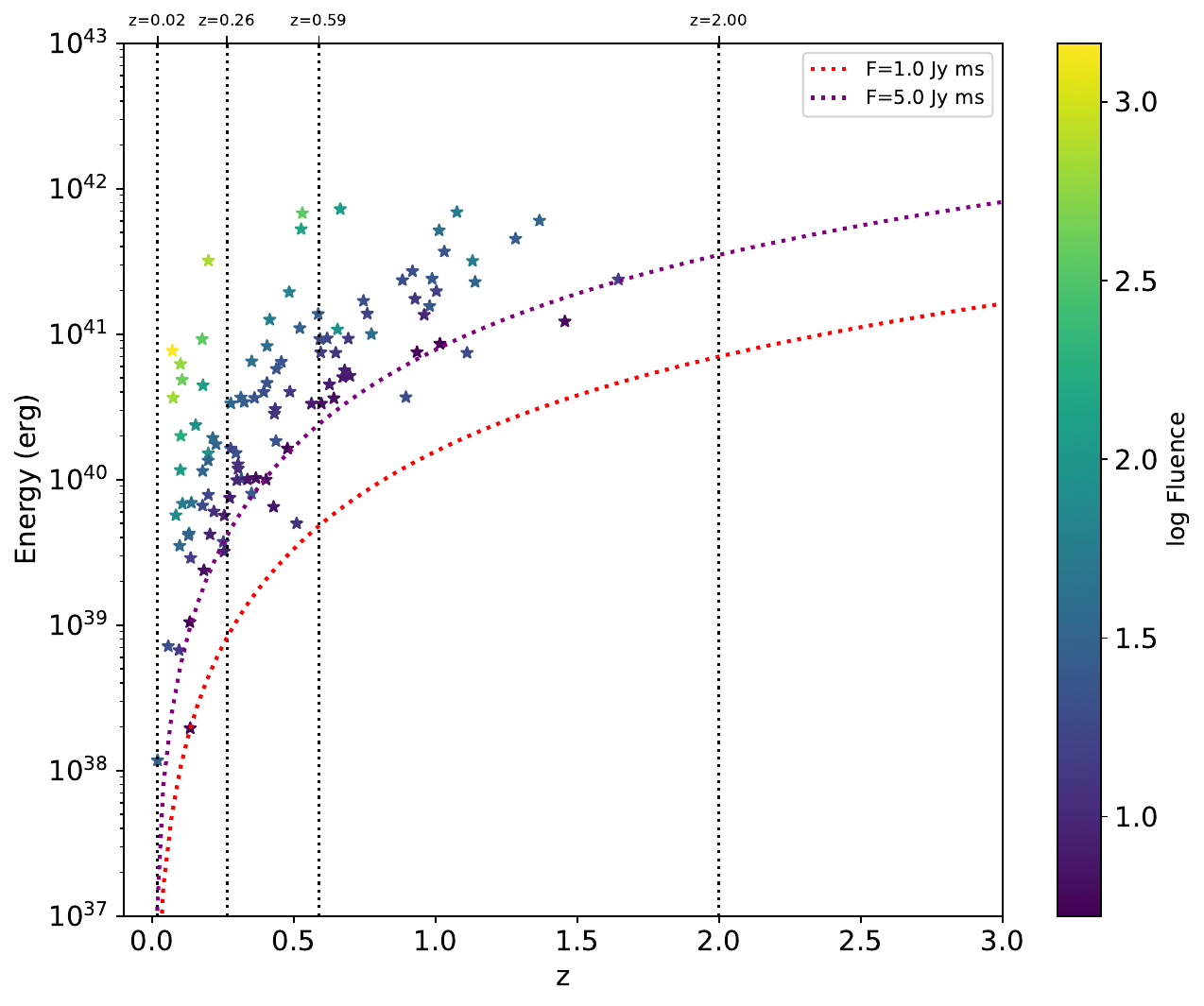}
  \caption{Isotropic-equivalent energy as a function of pseudo-redshift for the FRB samples. The redshift values were derived using the Bayesian method, while the energy values were calculated using Eq. \ref{eq:07}. The colour of each star corresponds to the observed fluence value, and the vertical dashed black lines denote the redshift bin boundaries: $z=0.02$, $0.26$, $0.59$, and $2.00$. The dashed red and purple lines represent the energy-redshift relations calculated assuming $F=1.0$, $5.0$ Jy ms and $\Delta{\nu}=400~\text{MHz}$.}
     \label{fig:energy}
\end{figure}

Significant morphological differences between repeaters and non-repeaters have been observed \citep{2021ApJ...923....1P, 2025ApJ...979..160S, 2025ApJ...982..158C}. As FRB emission mechanisms remain poorly understood, we focused on estimating the isotropic-equivalent energy. Typically, the FRB isotropic-equivalent energy is calculated as \citep{2018MNRAS.480.4211M, 2023ApJ...944..105S, 2024Univ...10..207Z, 2025ApJ...989..130B}
During the preparation of your manuscript for typesetting, the queries listed below have arisen.
Please answer each of these queries by adding the responses in this table, or by marking the corresponding corrections directly on this PDF file, or by writing your answers in the dedicated space on our platform, with your potential proofs corrections.

\begin{equation}
    E = \frac{4\pi d_\mathrm{L}^2(z)F\Delta\nu}{{(1+z)}^{2+\alpha}},
\label{eq:07}
\end{equation}

\noindent where $z$ is the redshift, $d_\mathrm{L}(z)$ is the luminosity distance at redshift $z$, $F$ is the observed fluence, $\Delta{\nu}$ is the frequency bandwidth of baseband data, and $\alpha$ is the spectral index from the assumed power-law spectrum $F(\nu)\propto \nu^{\alpha}$. The k correction, ${(1+z)}^{-\alpha}$, accounts for the fact that the frequency we observe differs from that at which it was emitted due to propagation \citep{2018MNRAS.480.4211M, 2023ApJ...944..105S, 2024Univ...10..207Z}. The constraint on $\alpha$ is not well established. \citet{2019ApJ...872L..19M} used data from 23 FRBs detected by the Australian Square Kilometre Array Pathfinder (ASKAP) to obtain a spectral index of $\alpha=-1.5^{+0.2}_{-0.3}$. In contrast, \citet{2022MNRAS.509.4775J} claimed that the index should be corrected to $\alpha = -0.65\pm0.3$, depending on whether all FRBs exhibit extremely narrowband emission. \citet{2023ApJ...944..105S} derive a spectral index of $\alpha=-1.39^{+0.86}_{-1.19}$. \citet{2025ApJ...982..158C} considered the correlations between properties and derive $\alpha = -2.29 \pm 0.29$ for a unified FRB population. Considering the intrinsic fluence spectral index and the complexity of the FRB spectrum, we adopt $\alpha=-1.5$ for energy calculation. Figure~\ref{fig:energy} shows the calculated isotropic-equivalent energy and redshift for each FRB, where the colour bar represents fluence. The dashed curves are fluence contours for $F = 1.0$ and $5.0$ Jy ms, assuming a 400 MHz bandwidth.

To facilitate an analysis of the volumetric FRB event rate evolution with redshift, the BaseCat 1 data are divided into three redshift bins, denoted as '\texttt{redshift bin 1}', '\texttt{redshift bin 2}', and '\texttt{redshift bin 3}'. Each bin containing an equal number of sources to minimise binning effects. The bins are defined by the redshift boundaries $z=0.02$, $0.26$, $0.59$, and $2.00$, indicated by vertical dashed black lines in Figure~\ref{fig:energy}.

%%%%%%%%%%%%%%%%%%%%%%%%%%%%%%%%%%%%%%%%%%%%%%%%%%%%%%%%%%%%%%

\section{Method}
\label{sec:method}

Investigating the energy function and event rate evolution with redshift provides critical constraints on FRB progenitors and populations. It depends on the FRB energy distribution and the number densities across redshifts. Typically, large samples observed from a single telescope are preferred, as their selection biases are more straightforward to consider, even when the number of events is limited. Here, we detail the construction of selection functions and the number density calculation method.

\subsection{Selection functions}
\label{subsec:sf}

Deriving selection functions is essential for correcting observational incompleteness. The selection functions are defined mathematically as the ratio of observed distributions ($P_\text{obs}(\xi)$) to intrinsic distributions ($P_\text{int}(\xi)$):

\begin{equation}
    s(\xi)=\frac{P_\text{obs}(\xi)}{P_\text{int}(\xi)},
    \label{eq:08}
\end{equation}

\noindent where $\xi$ represents an observable parameter. The observed distributions are determined by both the intrinsic FRB population properties and telescope-dependent selection effects.

The intrinsic distributions of FRB properties may become coupled in high-dimensional phase space because of complex observational biases and propagation effects. The CHIME/FRB Collaboration developed an injection system to characterise the full detection capability of CHIME, providing a basis for revealing the intrinsic distributions \citep{2023AJ....165..152M}. They investigated the selection functions available for CHIME, and derived the intrinsic distributions based on the CHIME/FRB Catalogue 1 data. Their results show that there are no strong correlations among intrinsic property distributions \citep{2023ApJ...944..105S, 2025ApJ...982..158C}. Furthermore, \citet{2023ApJ...944..105S}, who used the intensity data and took the signal-to-noise ratio (S/N) as a proxy of fluence, propose a joint S/N-DM distribution to investigate the FRB energy function and redshift distribution. This approach assumes that the  event rate of the FRB population evolved with the star formation rate characterised by parameter $n$, thereby introducing a prior cosmological evolution into the FRB population model.

However, we did not construct a joint fluence-DM distribution within our framework, because such a construction requires assumptions about the intrinsic energy and redshift distributions that are not independently established for the BaseCat 1 sample. To avoid introducing additional model dependence, we adopted the $V_{\rm max}$ method to estimate the volumetric event density. As a non-parametric approach, the $V_{\rm max}$ method does not require any prior assumption about the energy function or its event rate evolution (see Sect.~\ref{subsec:rate}). In other words, the posterior redshift and energy distributions cannot be used to construct the prior intrinsic fluence-DM distributions. Therefore, given the limited sample size and the absence of well-established intrinsic FRB distributions, we adopted a practical approximation by modelling the selection function as $s({\rm DM})s(F)$ instead of a joint selection function. This approximation is a compromise between modelling feasibility and the unknown nature of the true intrinsic distributions.

Systematic parameter differences between Catalogue 1 and BaseCat 1 suggest that a recalibration of selection functions for baseband data is necessary. We applied Anderson--Darling (A-D), Kolmogorov--Smirnov (K-S), and Mann--Whitney--Wilcoxon (M-W-W) statistical tests to investigate whether Catalogue 1 and BaseCat 1 observed samples are drawn from the same population, where a $p$ value < 0.05 indicates that the two samples are drawn from different underlying distributions with confidence levels > 95\% \citep{2020sdmm.book.....I}. Tests on G.L., G.B., DM, fluence ($F$), bandwidth ($\Delta\nu$), scattering time ($\tau_\text{scat}$), and duration ($w$) were performed. The resulting $p$ values are listed in Table~\ref{tab:p-value}. The results indicate that the sky coordinates and DM have identical observed distributions, whereas the other parameters do not. In other words, BaseCat 1 data maintain the same spatial and distance characteristics as Catalogue 1. Since the observed DM distribution is consistent between Catalogue 1 and BaseCat 1, we assume that the intrinsic DM distribution derived from Catalogue 1 can be applied to BaseCat 1. The fluence and burst bandwidth are related to the energy distribution, and the intrinsic fluence distribution must be reassessed due to significant changes in measured values of fluence and bandwidth.

\begin{figure*}
    \centering
    \includegraphics[width=\hsize]{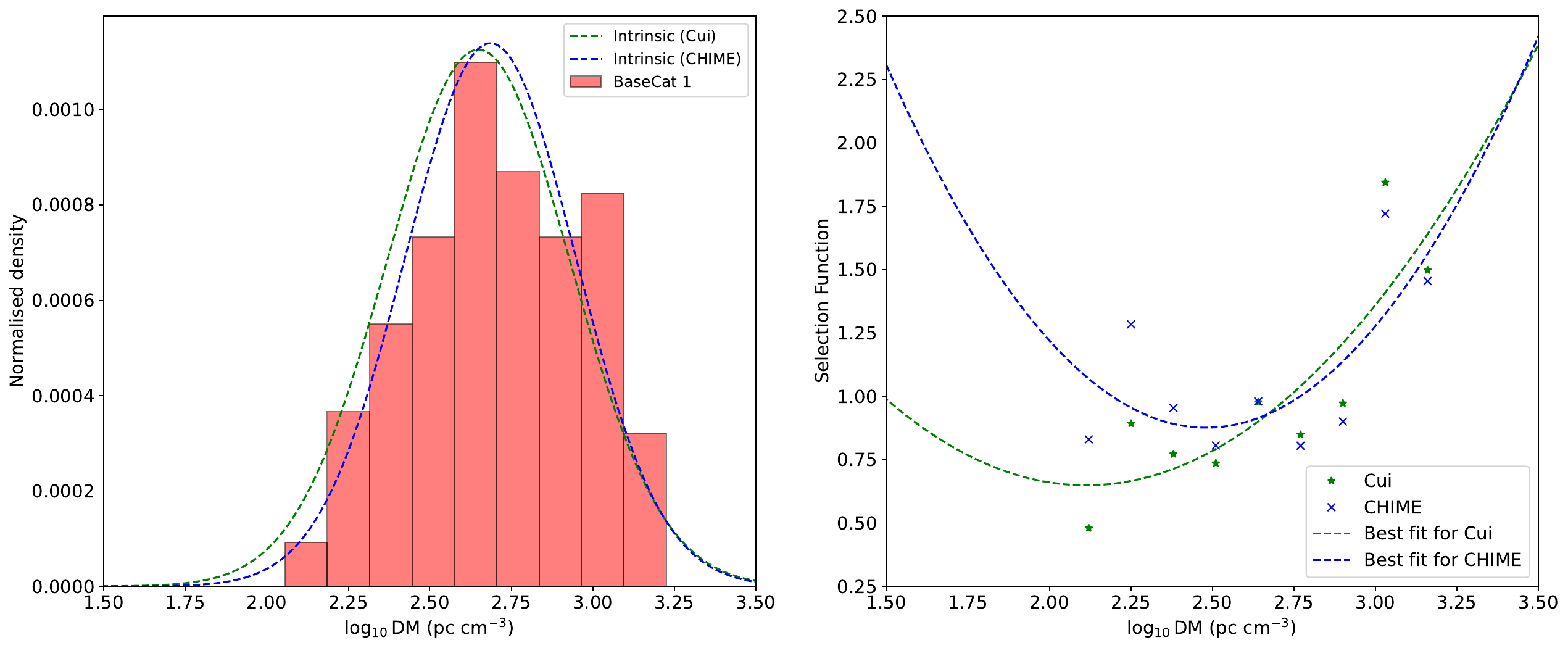}
    \includegraphics[width=\hsize]{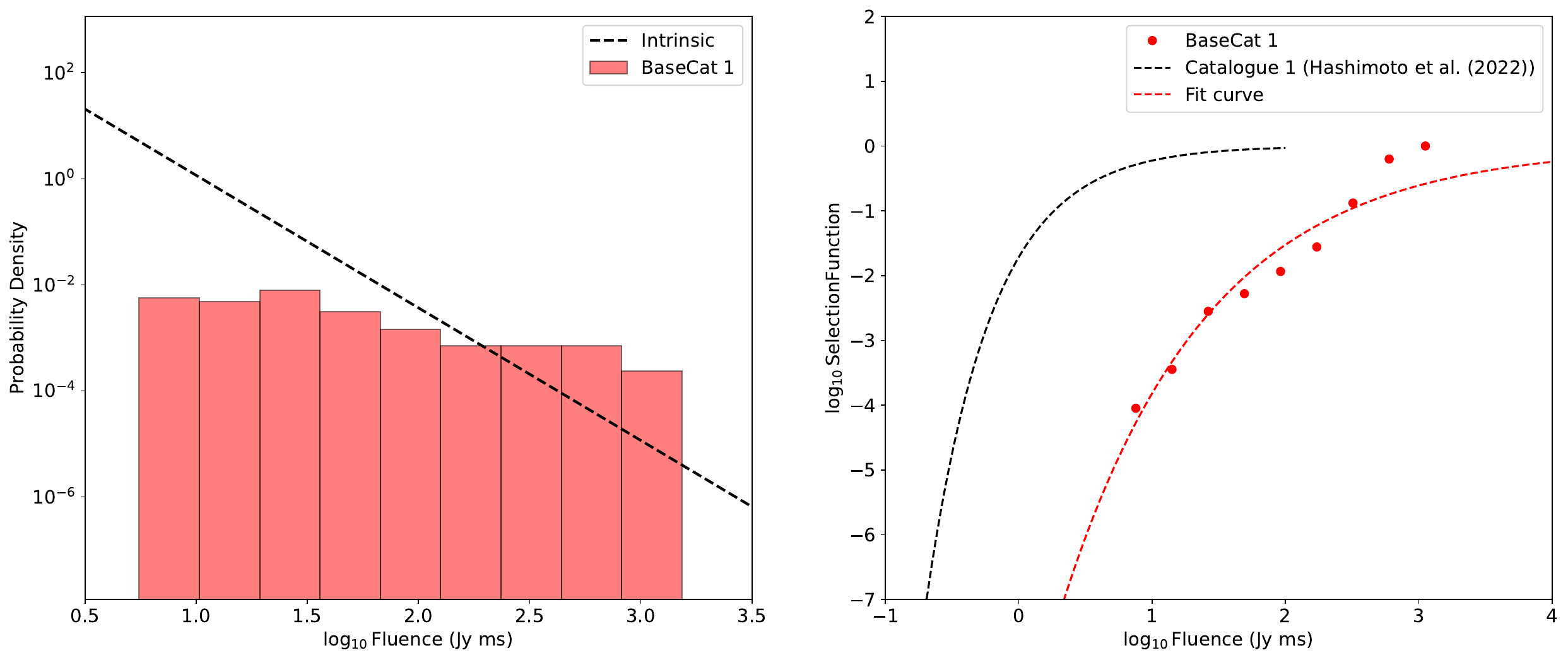}
    \caption{Left panels: Observed (red histogram) and intrinsic (dashed lines) distributions of DM (top) and fluence (bottom). The intrinsic distributions from \citet{2021ApJS..257...59C} and \citet{2025ApJ...982..158C} are shown with the dashed blue and green lines for comparison, while the theoretical intrinsic fluence distribution is indicated by the dashed black line. All distributions are normalised. Right panels: Selection functions for DM (top) and fluence (bottom). The DM selection function data points, compared with different intrinsic distributions, are plotted as green and blue markers. The fluence selection function data points are represented as red dots. The best-fit functions are demonstrated as dashed lines in the corresponding colour. The fluence selection function for Catalogue 1 derived by \citet{2022MNRAS.511.1961H} is represented as a dashed black line for reference.}
    \label{fig:Selection Function}
\end{figure*}

Determining intrinsic distributions presents significant methodological challenges in FRB population studies. The intrinsic DM distribution is assumed to follow a lognormal distribution:

\begin{equation}
    P_\text{int}(\text{DM})=\frac{1}{\sqrt{2\pi}\sigma (\text{DM}/\mu)}\mathrm{e}^{-\frac{\text{ln}^2(\text{DM}/\mu)}{2\sigma^2}},
    \label{eq:09}
\end{equation}

\noindent where $\mu$ and $\sigma$ are the scale and shape parameters, respectively. The CHIME/FRB Collaboration derived intrinsic distributions using the CHIME/FRB injection system \citep{2021ApJS..257...59C}. Later, \citet{2025ApJ...982..158C} considered the correlation between parameters and derived selection functions using the bias correction method. The intrinsic DM distributions from these works are shown as dashed green and blue lines in Figure~\ref{fig:Selection Function} \footnote{The intrinsic distribution data are kindly provided by Xiang-Han Cui.}. We adopted the intrinsic DM distribution parameters fitted by \citet{2025ApJ...982..158C} ($\mu=e^{6.52}=676.35$ and $\sigma=0.646$).

\begin{table}[!ht]
\caption{\label{tab:p-value}$p$ values of the parameters.}
\centering
\begin{tabular}{lccc}
    \hline\hline
    Parameters & $p_\text{A-D}$ & $p_\text{K-S}$ & $p_\text{M-W-W}$ \\
    \hline
    GL & $0.739$ & $0.863$ & $0.793$ \\
    GB & $0.231$ & $0.141$ & $0.794$ \\
    DM & $0.504$ & $0.666$ & $0.361$ \\
    Fluence ($F$) & $9.99\times10^{-5}$ & $2.73\times10^{-14}$ & $3.12\times10^{-18}$ \\
    Bandwidth ($\Delta\nu$) & $9.99\times10^{-5}$ & $3.21\times10^{-19}$ & $0.015$ \\
    Scattering time ($\tau_\text{scat}$) & $0.003$ & $0.012$ & $0.007$ \\
    Duration ($w$) & $9.99\times10^{-5}$ & $6.35\times10^{-20}$ & $5.39\times10^{-24}$ \\
    \hline
\end{tabular}
\tablefoot{The specific measured values for each property are adopted.}
\end{table}

Previous analyses used the S/N detection criterion as a proxy to study the fluence \citep{2022MNRAS.511.1961H, 2023ApJ...944..105S}, because the fluence measurements in Catalogue 1 have significant uncertainties from localisation and beam modelling. Typically, the standard selection methodology for CHIME/FRB Catalogue 1 data utilises well-validated S/N thresholds of 10 or 12 to ensure reliable event identification while maintaining sufficient sample sizes \citep{2022MNRAS.511.1961H, 2023ApJ...944..105S, 2025ApJ...982..158C}. However, the S/N calculation method for Catalogue 1 and BaseCat 1 is different. The S/N values of Catalogue 1 were obtained from the \texttt{fitburst} least-squares fitting program, and the BaseCat 1 values were calculated from the main peak of each pulse profile \citep{2024ApJ...969..145C}. Since the CHIME/FRB experiment stores baseband data for all events with an S/N > 12, and the fluence measurements from BaseCat 1 are significantly more accurate, we adopted fluence rather than S/N as the detection criterion. A power-law distribution is adopted for the intrinsic fluence distribution:

\begin{equation}
    P(F)\propto(\frac{F}{F_\text{limit}})^{\alpha_F-1},
    \label{eq:10}
\end{equation}

\noindent where $\alpha_F$ is the cumulative power-law index and $F_\text{limit}$ is the pivot fluence. \citet{2021ApJS..257...59C} obtain a power-law index of $-1.41\pm0.11(\text{stat.})^{+0.06}_{-0.09}(\text{sys.})$ from CHIME/FRB Catalogue 1 data, consistent with the theoretical integral Euclidean index of $\alpha_F=-1.5$. Unfortunately, the baseband data lack a corresponding spectral index analysis. We conservatively adopted the Euclidean expectation of $\alpha_F = -1.5$ (equivalent to a differential index of $-2.5$) and a reference fluence $F_\text{limit}=5.0$ Jy ms for analytical rigour. This choice aligns with both theoretical expectations and prior observations. The potential measurement uncertainties are discussed in Sect.~\ref{subsec:effects}. The observed distribution and intrinsic distribution of fluence are demonstrated in Figure~\ref{fig:Selection Function}.

The selection functions were calculated using Eq. (\ref{eq:08}). Unlike those of the CHIME/FRB injection system, the selection functions derived here were used for scaling weights when calculating the number density of FRBs discussed in Sect.~\ref{subsec:rate}. The DM selection function was fitted with a polynomial, while the fluence selection function follows the form given by \citet{2022MNRAS.511.1961H}

\begin{equation}
    \text{log}~s(F) = -a\times\text{e}^{b\times\text{log}~F},
    \label{eq:11}
\end{equation}

\noindent where $a$ and $b$ are fitting parameters. We assume that the FRB detection is complete at high fluence, so that $\text{log}~s(F)$ approaches zero. The best-fitting functions for DM and fluence are given as

\begin{equation}
    s(\text{DM}) = 0.904{(\text{log DM})}^2-3.818(\text{log DM})+4.682
    \label{eq:12}
\end{equation}

\noindent and

\begin{equation}
    \text{log}~s(F) = -9.564\times\text{e}^{-0.918~\text{log}~F}.
    \label{eq:13}
\end{equation}

\noindent The selection function for fluences below 5 Jy ms is unavailable because the observational completeness is unknown.

\subsection{$V_\mathrm{max}$ method and number density calculation}
\label{subsec:rate}

The $V_\mathrm{max}$ method \citep{1968ApJ...151..393S, 1980ApJ...235..694A} was used to derive the energy function. The energy function represents the number density of FRBs as a function of energy, which reveals the population distribution at a given distance. The $V_\text{max}$ method derives the energy functions without any prior assumption about their distributions. We followed the approach of \citet{2020MNRAS.494.2886H, 2020MNRAS.498.3927H, 2022MNRAS.511.1961H}, modifying only the redshift bin settings and selection functions, as described in Sects.~\ref{subsec:energy} and \ref{subsec:sf}.

The value $V_\mathrm{max}$ is the maximum comoving volume within which a source would be detectable above a given detection threshold. It is defined as

\begin{equation}
    V_\mathrm{max} = \frac{4\pi}{3}[d_{c,\mathrm{max}}^3(z)-d_{c,\mathrm{min}}^3(z)].
\label{eq:14}
\end{equation}

\noindent Here, $d_{c,\mathrm{min}}$ is the comoving distance to the lower bound of the redshift bin and is the maximum detectable comoving distance at which an FRB of a given energy is detectable by CHIME, corresponding to an upper-limit redshift $z_\mathrm{max}$. Beyond this distance, the signal becomes too faint due to decreasing S/N and fluence. The value of $d_{c,\mathrm{max}}$ for each FRB is determined by

\begin{equation}
    \frac{(1+z)E}{4\pi d_{\mathrm{L, max}}^2(z)\Delta\nu}\ge F_\mathrm{limit},
\label{eq:15}
\end{equation}

\noindent where $z$ is the pseudo redshift of the FRB, $E$ is the isotropic-equivalent energy, and $F_\text{limit}$ = 5 Jy ms is the fluence threshold. Moreover, $d_{\mathrm{L, max}} = (1+z) d_{c,\mathrm{max}}$ is the maximum luminosity distance. Note that $d_{c,\mathrm{max}}$ is constrained by the upper boundary $z_\mathrm{max}$ of its redshift bin. When Eq. (\ref{eq:15}) yields a $d_{c, \text{max}}$ exceeding this boundary, we set $d_{c, \text{max}}$ to the maximum comoving distance corresponding to the upper bound of the redshift bin.

The number density of each FRB per unit volume per unit time is calculated as \citep{2022MNRAS.511.1961H}

\begin{equation} 
    \rho_{i} = \frac{1+z_{\mathrm{FRB},i}}{V_{\mathrm{max}, i}\Omega_\mathrm{sky}t_\mathrm{obs}}, 
\label{eq:16} 
\end{equation}

\noindent where subscript $i$ denotes the $i$-th source, the factor $1+z_{\mathrm{FRB}}$ is the time conversion factor ($t_\mathrm{obs} = (1+z) t_\mathrm{rest}$), $\Omega_\mathrm{sky} \approx 250/41252.96 = 0.006$ is the sky coverage for the CHIME telescope \citep{2021ApJS..257...59C}, and the equivalent survey time for the CHIME/FRB baseband observation epoch $t_\mathrm{obs} = 0.18~\mathrm{yr}$, estimated following \citet{2025ApJ...986..100G}.

The number density calculated from Eq. (\ref{eq:16}) is not the true density due to the observational incompleteness, which is corrected by the selection functions as follows:

\begin{align} 
    \rho_{i,\mathrm{corr}} &= \rho_i \times W \times w_i(\mathrm{DM}) w_i(F) \notag \\
    &= \rho_i \times \frac{N_\mathrm{FRB}/\epsilon}{\Sigma_{i=1}^n w_i(\mathrm{DM}) w_i(F)} \times w_i(\mathrm{DM}) w_i(F),
\label{eq:17}
\end{align} 

\noindent where $w_i(\mathrm{DM})$ and $w_i(F)$ are the weight functions of DM and fluence, defined as the reciprocals of the selection functions (e.g. Eqs. (\ref{eq:12}) and (\ref{eq:13})). Here, $W$ is the scaling factor for the weight functions, determined such that the sum of weights over the selected samples equals $N_{\mathrm{FRB}}/\epsilon$. Moreover, $N_\mathrm{FRB}$ is the sample size, and $\epsilon = 39638/84697$ is the detection efficiency from the CHIME/FRB injection test \citep{2021ApJS..257...59C}.

The energy function $\Phi(z, E)$ is computed by summing the corrected density $\rho_\text{corr}$ for all sources within each redshift and logarithmic energy bin:

\begin{equation}
    \Phi_j(z, E) = \frac{\Sigma_k \rho_{j,k,\mathrm{corr}}(E)}{\Delta\mathrm{log~}E_j},
\label{eq:18} 
\end{equation}

\noindent where the subscript $j$ denotes the $j$-th energy bin, $k$ indicates the $k$-th FRB in the $j$-th energy bin, and $\Delta\mathrm{log}~E$ is the logarithmic energy bin width. We performed 10,000 Monte Carlo (MC) simulations, considering fluence measurement errors, to estimate the uncertainty in the energy function. The simulation procedure involves:

\begin{enumerate}
    \item Generating 10,000 redshift and fluence samples from the data distribution.
    \item Calculating the energy for each sample using Eq. (\ref{eq:07}). The simulated redshift, fluence, and measured bandwidth are used for each sample. The lower and upper energy errors are calculated from the 16th and 84th percentiles.
    \item Calculating $V_\text{max}$ and number density $\rho$ for each sample. The redshift bins are fixed for each simulation. The simulated samples are re-binned into each redshift bin.
    \item Calculating the selection function value and the weighted number density $\rho_\text{corr}$ for each simulated sample.
    \item Calculating the energy bins, then calculating $\Phi_j$ within each energy and redshift bin.
\end{enumerate}

%%%%%%%%%%%%%%%%%%%%%%%%%%%%%%%%%%%%%%%%%%%%%%%%%%%%%%%%%%%%%%

\section{Results}
\label{sec:results}

\subsection{Energy functions}

\begin{figure}[!ht]
\centering
\includegraphics[width=\hsize]{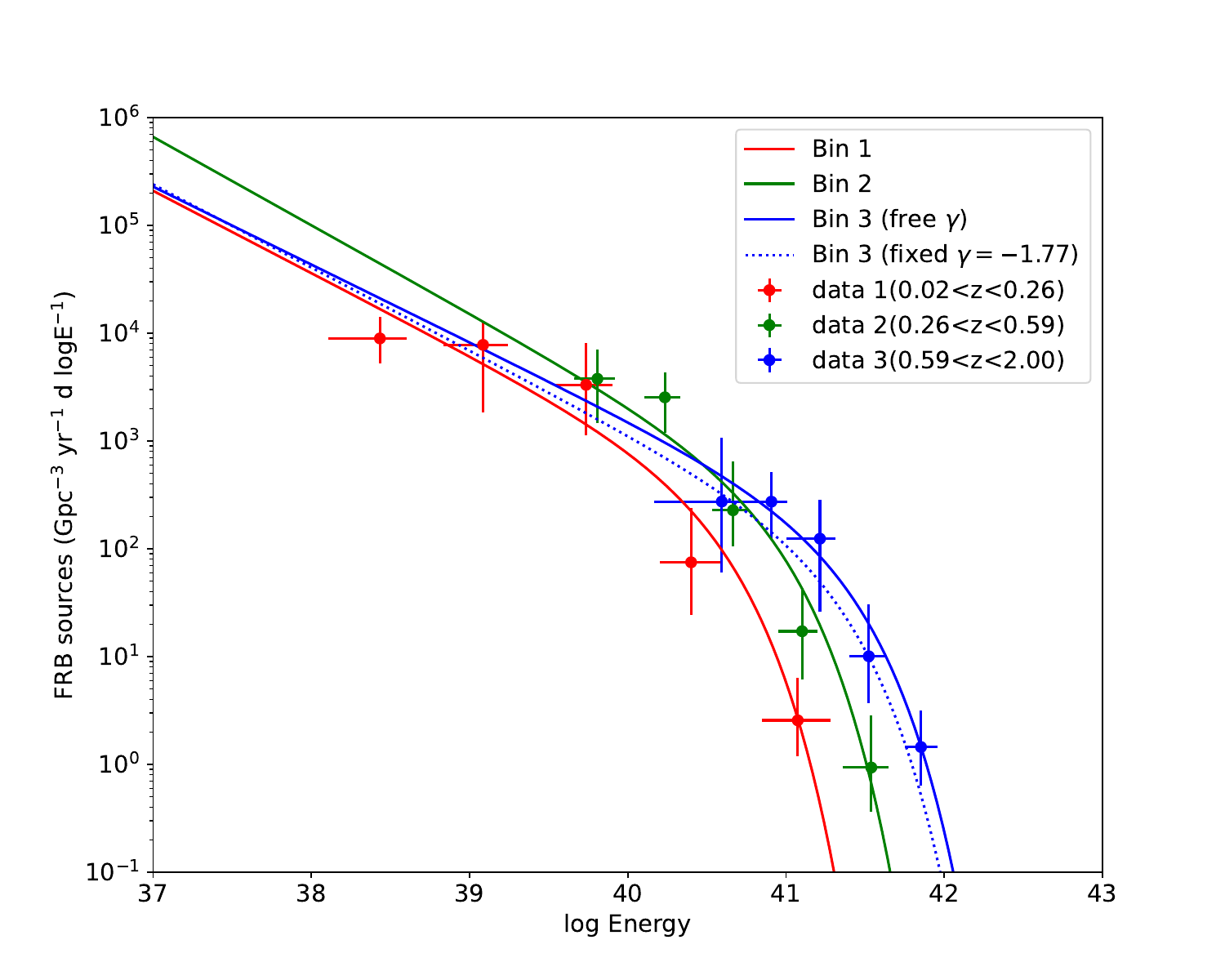}
  \caption{Energy function data for redshift bin 1 (red dots), bin 2 (green dots), and bin 3 (blue dots). The best-estimated energy function of each redshift bin is plotted as a dashed line, with the same colour as the energy function data.}
     \label{fig:energy_function}
\end{figure}

The energy function $\phi(E)$ describes the event rate per unit volume per unit energy within a redshift bin. The derived energy function data with errors within each redshift bin are shown in Figure~\ref{fig:energy_function}. The $\pm$1$\sigma$ uncertainties of energy functions were derived from the 84th and 16th percentiles of the MC simulation distributions, as detailed in Sect.~\ref{subsec:rate}. The rate declines slowly at lower energies and more steeply at higher energies, which indicates a much greater abundance of lower-energy events. The Schechter function is adopted to fit the data \citep{2022MNRAS.511.1961H, 2023ApJ...944..105S}:

\begin{equation}
    \phi(E)\text{d}E = \phi^{\star}{\left (\frac{E}{E^{\star}}\right )}^{\gamma}e^{-\frac{E}{E^{\star}}}\text{d}\left (\frac{E}{E^{\star}}\right),
\label{eq:19}
\end{equation}

\noindent where $\phi^{\star}$ is the normalisation factor, $E^{\star}$ is the cut-off energy, and $\gamma$ is the slope index. The logarithmic form of the energy function is commonly used:

\begin{equation}
    \phi(\mathrm{log}~E)\text{d}~\mathrm{log}~E = \phi^{\star}{\left (\frac{E}{E^{\star}}\right )}^{\gamma+1}e^{-\frac{E}{E^{\star}}}\text{d}~\mathrm{log}~E.
\label{eq:20}
\end{equation}

\noindent The Python package \texttt{bilby} \footnote{\url{https://git.ligo.org/lscsoft/bilby}} was utilised to estimate the best-fit parameters \citep{2019ApJS..241...27A}. \texttt{bilby} is a user-friendly Bayesian inference library that includes a built-in sampler \texttt{Dynesty} \citep{2019S&C....29..891H}. \texttt{Dynesty} is a pure-Python, dynamic nested sampling package for estimating Bayesian posteriors and evidence. It has been proven to be both fast and accurate. We set 30,000 live points for the \texttt{Dynesty} samplers, and the priors are set as: log $E^{\star}\sim \text{N}(41.0, 1.0)$, log $\phi^{\star}\sim\text{N}(3.0, 2.0)$, and $\gamma\sim\text{N}(-1.7, 0.5)$. The best-fit curves are shown with matching colours in Figure~\ref{fig:energy_function}. All estimated parameters are summarised in Table~\ref{tab:parameters}. The cut-off energy ($E^\star$) increases with redshift, which is consistent with previous studies \citep{2022MNRAS.511.1961H, 2023Univ....9..251Z}. The derived slope index of $\gamma = -1.64^{+0.42}_{-0.44}$ within \texttt{redshift bin 1} is steeper than the value of $\gamma \approx -1.4$ reported by \citet{2022MNRAS.511.1961H} using Catalogue 1 data.

Because of missing low-energy data points, we fixed the slope index of \texttt{redshift bin 3} to the \texttt{redshift bin 2} value and fit the other parameters. The best-fit curve is demonstrated as a dashed blue curve in Figure~\ref{fig:energy_function}, and the parameters are shown in Table~\ref{tab:parameters}. The results show no significant difference between the two methods.

\begin{table}[!ht]
\caption{\label{tab:parameters}Parameters of the energy function.}
\centering
\begin{tabular}{cccc}
    \hline\hline
    Redshift bin & $\mathrm{log}~\phi^{\star}$ & $\mathrm{log}~E^{\star}$ & $\gamma$ \\
    \hline
    $0.02<z<0.26$ & $2.55^{+0.94}_{-1.61}$ & $40.33^{+0.54}_{-0.50}$ & $-1.64^{+0.42}_{-0.44}$ \\
    $0.26<z<0.59$ & $2.79^{+0.53}_{-0.60}$ & $40.63^{+0.26}_{-0.28}$ & $-1.77^{+0.29}_{-0.28}$ \\
    $0.59<z<2.00$ (free $\gamma$) & $2.20^{+0.50}_{-0.51}$ & $41.15^{+0.22}_{-0.47}$ & $-1.68^{+0.28}_{-0.28}$ \\
    $0.59<z<2.00$ (fixed $\gamma$) & $2.14^{+0.44}_{-0.45}$ & $41.21^{+0.18}_{-0.48}$ & $-1.77$\tablefootmark{a}
    \\
    \hline
\end{tabular}
\tablefoot{The slope index for redshift bin 3 is fixed to the value of redshift bin 2.}\\
\end{table}

\subsection{Rate evolution as a function of redshift}

The FRB volumetric event rate density, in units of events per cubic gigaparsec per year, is derived by integrating the energy function from a threshold energy $E_\text{lim}$ to infinity as

\begin{equation}
    R(>E_\text{lim}) = \int_{\text{log}~E_\text{lim}}^{\infty}\phi^\star\left(\frac{E}{E^\star}\right)^{\gamma+1}e^{-E/E^\star}~\text{d}~\text{log}E.
\label{eq:21}
\end{equation}

\begin{figure}[!ht]
\centering
\includegraphics[width=\hsize]{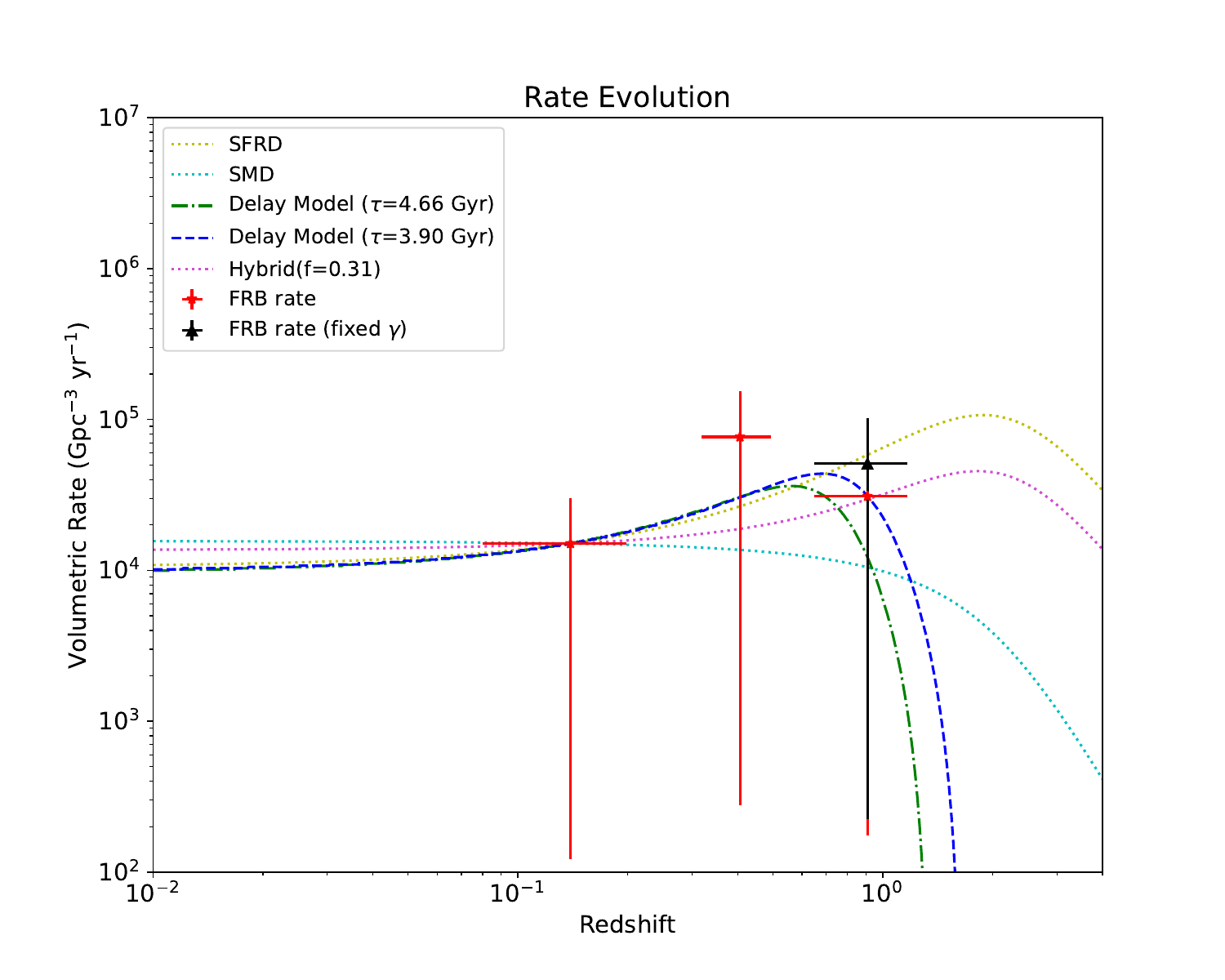}
  \caption{FRB volumetric event rate evolution as a function of redshift. The red dots with error bars represent the volumetric event rates within each redshift bin. The event rate within \texttt{redshift bin 3} with fixed $\gamma$ is illustrated as a black dot. The cosmic SMD evolution \citep{2014ARA&A..52..415M, 2018A&A...615A..27L} and the cosmic SFRD evolution \citep{2017ApJ...840...39M} are shown as dashed yellow and cyan lines, respectively. The hybrid model with $f=0.31$ is illustrated as a dashed purple line \citep{2025ApJ...986..100G}. The best-fit of the delay time model is plotted as dashed green and dashed blue lines for the normal method and the fixed $\gamma$ method, respectively.}
     \label{fig:evolution}
\end{figure}

\noindent The determination of the FRB threshold energy is arbitrary. The typical burst energy of FRBs ranges from $10^{38}$ to $10^{42}$ erg. We set the threshold energy as $E_\text{lim}=10^{38}~\text{erg}$ and integrate the energy function over the logarithmic energy range to calculate the event rate density within each redshift bin. We perform 10,000 MC simulations to derive the event rate. The integrated FRB event rates are demonstrated as red dots with errors in Figure~\ref{fig:evolution}. The redshift values and errors are given by the median and standard deviation of sources within each redshift bin, respectively. The event rate uncertainties were adopted from 1$\sigma$ Poisson errors, to which we added the fitting uncertainties of the energy function. The rate obtained by fixing $\gamma$ in redshift bin 3 is illustrated as a black dot. The volumetric rate error is large due to the limited source number and the absence of low-energy events.

The evolution of the FRB population with redshift is considered. If FRBs primarily arise from young stellar populations such as core-collapse supernovae (CCSNe), their rate evolution may follow the cosmic star formation rate density (SFRD). The SFRD evolution function is adopted from \citet{2017ApJ...840...39M}:

\begin{equation}
    \dot{m}_{*}(z) = \frac{0.01(1+z)^{2.60}}{1+[(1+z)/3.2]^{6.20}},
    \label{eq:22}
\end{equation}

\noindent which reproduces the observed SFRD better at $z < 10$. Conversely, if most FRBs originate from older stellar populations (such as neutron stars), their evolution would follow the cosmic stellar-mass density (SMD). The SMD evolution is defined as \citep{2014ARA&A..52..415M}

\begin{equation}
    m_*(z) = (1-R)\int_z^\infty\frac{\dot{m}_*(z')}{(1+z')H(z')}\text{d}z',
    \label{eq:23}
\end{equation}

\noindent where $R=0.27$ is the 'return' fraction for the Salpeter initial mass function (IMF), meaning the mass fraction of each generation of massive stars that is returned to the ISM and IGM after stellar death \citep{2014ARA&A..52..415M}. Moreover, $H(z)$ is the Hubble parameter:

\begin{equation}
    H(z) = H_0\sqrt{\Omega_{m}{(1+z)}^3+\Omega_{\Lambda}},
    \label{eq:24}
\end{equation}

\noindent where $H_0$ is the Hubble constant and $\Omega_m$ and $\Omega_\Lambda$ are the matter and dark energy density parameters, respectively.

Fast radio bursts may originate from binary systems \citep{2021ApJ...907..111Z, 2022NatCo..13.4382W, 2023NatAs...7..579M}. Such binary systems must undergo a long inspiral phase before the final coalescence; therefore, an additional delay is expected between the formation of the progenitor system and the FRB bursts. Thus, the evolutionary trend of FRBs produced from binary systems would follow the star formation activity but exhibit a constant delay time $\tau$. This delay can be characterised by the model:

\begin{equation}
    \phi_*(z,\tau) = \frac{\dot{m}_*(Z(t_\text{lb}(z)+\tau))}{\dot{m}_*(z=0)},
\label{eq:25}
\end{equation}

\noindent where $t_\text{lb}(z)$ is the look-back time at redshift $z$, and $Z(t_\text{lb}(z)+\tau)$ is a function for converting a look-back time to redshift. The parameter $\tau$ represents the average delay time between the emergence of the FRB population and star formation. We find a best-fit delay time of $\tau=4.66$ Gyr for the no fixed $\gamma$ scenario and $\tau=3.90$ Gyr for the fixed $\gamma$ scenario using the \texttt{scipy.optimize.least\_squares} package. The corresponding fit curves are shown as dashed green and blue lines in Figure~\ref{fig:evolution}.

Repeaters and non-repeaters may originate from either distinct populations or different evolutionary epochs of binary systems. An alternative possibility is that FRB populations track two distinct evolution patterns in some proportion. A hybrid model has been proposed \citep{2023ApJ...944..105S, 2025ApJ...986..100G}:

\begin{equation}
    \phi(z, f) \propto f\times\dot{m}_*(z) + (1-f)\times m_*(z),
    \label{eq:26}
\end{equation}

\noindent where $f$ is the fraction of FRBs that are produced by the young stellar population. \citet{2025ApJ...986..100G} concluded that $f=0.31$ can account for the observations from CHIME/FRB Catalogue 1 and BaseCat 1 data.

The evolution trends are shown in Figure~\ref{fig:evolution}. These trends are arbitrarily scaled to match the volumetric rate at the median of \texttt{redshift bin 1} for comparison. We calculated reduced $\chi^2$ values to quantify the difference between the FRB event rate evolution trend and the cosmic evolution models. By calculating the sum of squared deviations weighted by uncertainties, the $\chi^2$ values are 0.22 and 0.13 for the SFRD and SMD scenarios. The corresponding $p$ values are 0.88 and 0.94, indicating that the null hypothesis is not rejected at the 1$\sigma$ level for both SFRD and SMD. The $\chi^2$ value of the best-fit delay model is 0.68, and the $p$ value is 0.51.

In summary, there is no significant evidence to rule out the possibility that the FRB population tracks SFRD or SMD evolution. A hybrid model is attractive for the FRB population. The consistent trend between the FRB event rate and the delay time model suggests that most FRBs could originate from delayed scenarios such as binary systems. The large uncertainty in FRB event rate calculation and the limited number of baseband sources result in substantial uncertainty in the delay time estimation; the delay time distribution is still not determined.

%%%%%%%%%%%%%%%%%%%%%%%%%%%%%%%%%%%%%%%%%%%%%%%%%%%%%%%%%%%%%%

\section{Discussion}
\label{sec:discussion}

Using CHIME/FRB BaseCat 1 data, we investigated how improved localisation precision and flux density or fluence measurements affect the FRB energy function and rate evolution. Our results differ from those reported by \citet{2022MNRAS.511.1961H} for several reasons. First, the BaseCat 1 data are more accurate than Catalogue 1 data. There are significant differences in observational parameters between the CHIME/FRB Catalogue 1 and BaseCat 1 data due to improved measurement precision. Second, the construction of selection functions is different. We adopted the intrinsic distribution derived by \citet{2025ApJ...982..158C}, and the theoretical spectral index was assumed. Additionally, we set redshift bins containing an equal number of sources. Despite being limited by insufficient data, our work reveals several key features. Our results provide a new view on FRB energy functions and rate evolution analysis, suggesting a possible FRB origin channel. In this section, we discuss the caveats of our results, analyse the robustness, and outline prospects for future data releases.

\subsection{FRB energy functions}

The energy function reflects different properties from the luminosity function. The isotropic-equivalent luminosity is calculated from burst flux density, whereas the isotropic-equivalent energy was calculated from the fluence of the burst. The flux density was calculated by measuring the pulse amplitude at the maximum of the pulse profile, while the fluence was calculated by integrating the flux density over the burst duration \citep{2021ApJS..257...59C, 2024ApJ...969..145C}. The fluence and flux density are associated with the burst profile, duration and emission mechanism. The durations of FRBs range from 0.1 to 10 ms according to the BaseCat 1 data, and the brighter bursts have shorter durations \citep{2025ApJ...979..160S}. The fluence measurement is less affected by the observational time resolution than the flux density \citep{2018MNRAS.474.1900M}. Due to these factors and the poorly understood nature of FRB emission and beaming, the distributions of luminosity and energy may differ. It is noted that the isotropic-equivalent energy distribution of the FRB population might differ from the true burst energy distribution due to the unclear emission properties of FRB radiation.

Studies of the FRB energy function are essential for understanding the FRB population. The true energy of FRBs depends on burst shape, spectral properties, bandwidth and emission mechanisms. Assuming a Schechter function, \citet{2020MNRAS.494..665L} reported a luminosity function slope of $-1.79^{+0.31}_{-0.35}$ using combined data from Parkes, ASKAP, Arecibo, GBT and UTMOST. \citet{2020MNRAS.494.2886H, 2020MNRAS.498.3927H} estimated the time-integrated luminosity functions and derive a best-fitting slope index of $-1.35^{+0.51}_{-0.50}$ under a power-law shape assumption. \citet{2022MNRAS.511.1961H} derive energy function slope indices of $-1.4^{+0.7}_{-0.5}$ and $-1.1^{+0.6}_{-0.4}$ for different redshift bin settings for apparently non-repeaters from the CHIME/FRB Catalogue 1 data. \citet{2023ApJ...944..105S} obtain a flatter energy function slope of $-1.3^{+0.7}_{-0.4}$. \citet{2025ApJ...986..100G} calibrated the 2D fluence-DM distribution using baseband data and find that the energy distribution is described well by a Schechter function with a slope of $-1.94_{-0.12}^{+0.14}$. Our results are consistent with \citet{2025ApJ...986..100G} but steeper than those of \citet{2022MNRAS.511.1961H} and \citet{2023ApJ...944..105S}.

Considering the possibility that repeaters and non-repeaters have different origins, the energy functions of the two populations might be different. Typically, the energy function of non-repeaters reveals the number density of sources with different energies, whereas the energy function of repeaters reveals the burst pattern and waiting time distribution \citep{2022Natur.609..685X, 2024NatAs...8..337K}. Therefore, the energy function we discuss is an ensemble average over the behaviour of all FRBs because the origins of the FRB populations remain unclear. However, there may be repeaters within the non-repeater population, which could introduce unknown biases. The mixed population may cause the energy function slope index $\gamma$ to evolve with redshift. \citet{2022MNRAS.511.1961H} fixed the slope index $\gamma$ of the energy function within different redshift bins. We discuss the possibility of parameter evolution by treating the slope index as a free parameter. The fixed slope index is also considered within the last redshift bin. As shown in Table~\ref{tab:parameters}, our results are steeper than those from \citet{2022MNRAS.511.1961H} and \citet{2023ApJ...944..105S} but consistent with \citet{2025ApJ...986..100G}. No evidence of slope index evolution is found from the limited data.

The sources with extremely low energies ($<10^{38}$ erg) were discarded due to their proximity. As shown in Eq. (\ref{eq:07}), the primary uncertainty in energy calculation arises from fluence and bandwidth measurements. The discrepancies among energy functions derived from previous CHIME/FRB data show the importance of energy calculation from more accurate observational parameters (fluence and bandwidth). The relative uncertainty of fluence from the telescope is about 10--13\% \citep{2024ApJ...969..145C}. The redshift estimation also affects energy calculation, and the redshift uncertainty stems from pseudo redshift estimation. An error analysis shows that the relative error in the estimated redshift decreases as redshift increases. This occurs because the error from the host galaxy DM contribution affects low DM FRBs more severely, highlighting the importance of localised FRBs and host galaxy studies.

Since bright bursts may be flagged as radio frequency interference or saturate receivers, FRB events from the Milky Way and nearby galaxies are rare \citep{2024NatAs...8..337K}. Furthermore, the different frequency coverage among Parkes (1157--1546 MHz), ASKAP (1129--1465 MHz) and CHIME (400--800 MHz) may lead to different energy function slopes \citep{2022MNRAS.510L..18J, 2023ApJ...944..105S}. Related research is outside the scope of this work.

\subsection{FRB event rate and evolution}

The number density of FRB sources was re-weighted based on selection functions. The FRB volumetric event rate evolves with redshift, which could help distinguish FRB populations. \citet{2022MNRAS.511.1961H} argued that FRBs may not track SFRD evolution, while \citet{2023ApJ...944..105S} find no correlation between the FRB event rate and the SMD rate. The FRB rate evolution in Figure~\ref{fig:evolution} is different from the SMD or SFRD trends. Neither the SFRD nor the SMD evolution scenario can be ruled out due to the limited baseband samples. \citet{2022ApJ...924L..14Z} propose a hybrid model, in which 80\% of the sources belong to the delayed component and 20\% belong to the SFRD components. They tested different delay time distributions and conclude that FRBs may evolve following large delay times ($\sim 10$ Gyr). \citet{2024ApJ...973L..54C} find that the rate of non-repeaters declines with redshift, similar to sGRBs. They conclude that neutron stars and black holes are closely related to the origins of FRBs. \citet{2025ApJ...986..100G} suggest that FRB sources may contribute from both young and old stellar populations. The proportion is 0.31 for SFRD components and the remainder for SMD components. They also claim that large delay times are excluded by the baseband data with good confidence.

Our results show that FRBs may track SFRD evolution with a delay time of $\sim$ 4--5 Gyr. We assume that FRBs are primarily produced from binary neutron star (BNS) or delayed CCSNe populations. The delay times associated with the non-CCSNe channel are expected to have extended delay time distributions of several gigayears \citep{2013A&A...558A..39T}, which is longer than the expected delay time of CCSNe from stellar-merger remnants (about 50--250 Myr) \citep{2024Natur.635...61S}. The FRB events from BNS mergers should trace SFRD evolution with a delay time of a few gigayears because of the complex interactions. If the FRB populations are indeed dominated by these progenitors, this would suggest that the delay time distribution shows a wide range for different sub-populations \citep{2022ApJ...924L..14Z}.

Cosmological FRBs have burst energies of $10^{38}$ to $10^{42}$ erg, which are several orders of magnitude higher than those of the galactic source FRB~20200428. The burst fluence of FRB~20200428 is as high as $10^5$ Jy ms, corresponding to the burst energy of $\sim10^{34}~\text{erg}$ \citep{2020Natur.587...54C, 2020Natur.587...59B}. The burst rate for sources with such energies is $\sim 10^8~\text{Gpc}^{-3}~\text{yr}^{-1}$. However, the rate of bursts with $E_\text{iso}>10^{34}~\text{erg}$ from galactic magnetars is quite low based on the detection from SGR~1935+2154 and nearby galaxies \citep{2020Natur.587...54C}. As a result, it remains unclear whether cosmological FRBs share an origin with these galactic events.

In any case, additional samples at $z\sim2$ are needed to test the true redshift evolution of the FRB population. However, detecting high-redshift FRBs is challenging. The $\mathrm{DM_{IGM}}$ component is approximately 1700 and 3800 $\mathrm{pc~cm^{-3}}$ for bursts at $z = 2$ and $z = 5$, respectively. Although the maximum DM limit in current telescopes is quite high (e.g. CHIME searches up to 13,000 $\mathrm{pc~cm^{-3}}$), bursts may be missed due to intra-channel dispersion smearing, reduction in peak S/N due to scatter broadening, and cosmological pulse-width dilation, which also reduces the peak S/N \citep{2025PASA...42...17H}. Typically, radio telescopes set a maximum DM limit for searching. Additionally, the precise localisation for high-redshift events is also challenging.

\subsection{Baseband data advantages and disadvantages}
\label{subsec:baseband}

Previous work using CHIME/FRB Catalogue 1 data may underestimate the true FRB burst energy for several reasons. First, the fluence measurement of Catalogue 1 is underestimated. Therefore, the S/N is adopted as a fluence proxy to study the energy distribution, which may not be reliable due to the unclear correlation between fluence and S/N. Second, the frequency bandwidth changes significantly. Most bursts from BaseCat 1 exhibit a broader frequency band than those from Catalogue 1. The burst emissions even fill the entire frequency band. Therefore, the isotropic-equivalent energy could be underestimated. The broadband bursts appear across the entire energy range and redshift range. One can conclude that broadband emission in the CHIME/FRB frequency band is common in the FRB population. Broadband emission leads to higher energy, which increases the number of higher-energy bursts. The energy function, consequently, may be shifted to higher energy, resulting in higher cut-off energy $E^\star$. If such a broadband emission scenario dominates the FRB population, the effect on the slope index $\gamma$ is negligible. Otherwise, $\gamma$ could evolve with energy or redshift.

The true energy is also influenced by the spectral index $\alpha$. The FRB spectral properties are relatively complex, and the interpretation of $\alpha$ requires further consideration. If the FRB emission bandwidth is similar to or larger than the telescope bandwidth, FRBs with broadband emission can be interpreted using the spectral index. $F\sim\nu^{\alpha}$ is applicable to an individual broadband FRB or the full FRB population. In contrast, if the FRB population rate is frequency-dependent, such an interpretation may not be suitable. In such a case, the rate follows $\Phi\sim\Phi(z)\nu^{\alpha}$, which directly affects FRB rate evolution \citep{2023ApJ...944..105S}. The frequency band and central frequency of emission also affect the determination of $\alpha$. Given the complicated spectrum, the CHIME/FRB Collaboration uses two parameters ($\gamma$ and $r$) to describe FRB spectral properties observed by CHIME/FRB \citep{2021ApJ...923....1P}:

\begin{equation}
    F(\nu)\propto(\nu/\nu_0)^{\gamma+r\text{ln}(\nu/\nu_0)},
    \label{eq:27}
\end{equation}

\noindent where $\nu_0=400.1953125~\text{MHz}$. As $r\xrightarrow{}0$, the function approaches a power law. The bandwidth and peak frequency are derived directly from this functional form, and the parameters ($\gamma$, $r$) would strongly affect energy calculation and spectrum analysis. Due to the broadband behaviour within a lower frequency band, $\alpha$ is interpreted as a frequency spectral index, which is also applied to individual broadband bursts \citep{2023ApJ...944..105S}. The ideal value $\alpha=-1.5$ is adopted to simplify the spectrum, and the model-dependence of $\alpha$ needs thorough treatment.

Large single-telescope samples simplify the procedure of population investigation, and precise fluence measurements improve constraints on the energy function. However, there are drawbacks to baseband data. The baseband data system only stores events above a certain real-time S/N cut-off (typically $10$--$12$), resulting in a lack of low-fluence events \citep{2024ApJ...969..145C}. Extending the lower limit of the observed energy range is important for constraining the energy function. The lowest-energy event detected from a far distance is affected by the detection sensitivity of the telescope. As shown in Figure~\ref{fig:energy}, the red curve represents a fluence of $1.0$ Jy ms, which is below the fluence limit used in our work (purple curve). Consequently, the derived energy function in higher redshift bins exhibits a narrower energy range, leading to poorer constraints. Greater detection sensitivity is crucial for detecting low-energy sources from far distances.

As a key distance indicator, DM reflects the distance distribution of FRBs. Due to fluctuations in the DM components, the pseudo redshift of low DM events could vary significantly. To minimise contamination from nearby sources, we set the redshift lower limit as $z=0.02$ and discard sources below it. The $\mathrm{DM_{host}}$ contribution requires better constraints from localised events to reduce systematic uncertainties in redshift estimation. Considering that $\mathrm{DM_{host}}$ may not be too large for the majority of the FRB population, sources with large $\mathrm{DM_{ex}}$ are less affected by $\mathrm{DM_{host}}$ uncertainties.

A bias towards high-DM events is expected in CHIME observations, since DM smearing dilutes the burst signal in time \citep{2021ApJS..257...59C}. Although the scattering timescale measurements of BaseCat 1 are more robust than those of Catalogue 1 due to finer time resolutions \citep{2024ApJ...969..145C}, it is still hard to investigate the effect on the DM distribution. This bias may affect the DM distribution and the measured burst width, although its relative effect is smaller for bursts with wider intrinsic profiles. From another perspective, if FRBs have flatter spectra, the effect of smearing is reduced since a large fraction of the signal would come from higher frequencies. This bias requires careful consideration in future datasets. Given the uncertainty from intrinsic scattering distribution, the effects of duration and scattering time on DM are ignored due to the weak correlations. Only parameters related to distance and energy distributions were considered. Besides, the CHIME/FRB injection system cannot model propagation and scattering effects \citep{2023AJ....165..152M}. The derivation of selection functions depends on both intrinsic distributions and observational distributions. We assumed independent intrinsic distributions to simplify the construction of 1D selection functions, which could introduce extra unquantified uncertainty. The correlation between fluence and DM is expected to affect our results slightly, and we tested that the 1D selection function is effective at correcting selection effects.

Given that the S/N of detected bursts increases as the square root of the bandwidth of the burst \citep{2003ApJ...596.1142C}, future radio telescopes with ultra-wideband (UWB) receivers, such as the Canadian Hydrogen Observatory and Radio-transient Detector (CHORD; observing between 300 and 1500 MHz; \citep{2019clrp.2020...28V}), the Deep Synoptic Array 2000 (DSA-2000; observing between 700 and 2000 MHz; \citep{2019BAAS...51g.255H}), and the Qitai radio telescope (QTT; observing between 700 and 4000 MHz with 15-cm UWB receivers; \citep{2023SCPMA..6689512W}), are expected to improve detection capabilities.

\subsection{Uncertainties from intrinsic distributions and selection functions}
\label{subsec:effects}

The observed FRB population is incomplete due to complex observational effects. The intrinsic distributions are key to understanding FRB population properties, and these distributions are difficult to determine. The intrinsic DM distribution exhibits a lognormal shape, which is common in cosmological distributions \citep{2021ApJS..257...59C, 2023ApJ...944..105S, 2025ApJ...982..158C}. The injection system reveals that there is no significant correlation between detection efficiency and DM, indicating that the rarity of high-DM events is possibly due to the faint flux of bursts \citep{2023AJ....165..152M}.

Whether the intrinsic fluence distribution of FRBs is a power-law distribution is unclear. Accurate determination of the intrinsic fluence distribution slope index $\alpha_{F}$ in Eq. (\ref{eq:10}) requires more accurate observations. \citet{2022MNRAS.509.4775J} find that $\alpha_F$ follows a Euclidean distribution with a higher fluence threshold and provide a range of $-1.7\le\alpha_F\le-1.3$. We conservatively adopted $\alpha_F=-1.5$ in this work, and the impact of different distribution slope indices is considered. Adopting different slopes for the fluence selection function changes the volumetric rates, because the number of sources with a given burst energy differs when the incompleteness is corrected with a different $\alpha_F$. Assuming a Gaussian distribution of the spectral index with $\mu=-1.5$ and $\sigma=0.3$, we find that the index change mainly affects $\phi^\star$ and does not significantly affect $E^\star$ and $\gamma$ values. Within this uncertainty, the volumetric event rate varies by up to a factor of 10. Thus, the event rates derived in our work might be different from the true rate of the FRB population. However, this uncertainty does not affect our conclusions regarding the slope index of the energy function or the event rate evolution.

It is noted that the selection functions are only available for specific datasets and a limited parameter space. Uncertainties remain in $w(\text{DM})w(F)$, and further consideration is required. The correction of event rate density is associated with the scaling factor $W$ as well. This scaling factor is used to re-scale each sample so as to match the data. The value $\epsilon$ represents the detection efficiency of CHIME during the observation, which is used to correct $\rho_{i, \text{corr}}$ globally and consequently increase the overall event rate density. The event rate density depends on the modelling method, and the true density is difficult to determine. Besides, the second-order polynomial is used for the DM selection function, and the function form of the fluence selection function is chosen empirically to reproduce the observed distribution. Fitting the selection functions also introduces uncertainties. Therefore, additional observational data are needed to conduct comprehensive analyses. Joint distributions --- particularly for pulse width and scattering time --- are needed to study intrinsic distributions and selection functions. The upcoming baseband data of CHIME/FRB Catalogue 2 and injection experiments will be helpful for FRB population investigation.

%%%%%%%%%%%%%%%%%%%%%%%%%%%%%%%%%%%%%%%%%%%%%%%%%%%%%%%%%%%%%%
\section{Conclusions}
\label{sec:conclusion}

The energy function and volumetric event rate evolution provide critical constraints on FRB progenitors, though previous studies have yielded varied results. Here, we used CHIME/FRB baseband data to derive these measurements. The BaseCat 1 data, published by the CHIME/FRB Collaboration, provide improved measurements of burst properties --- such as localisation, bandwidth, flux density, and fluence --- compared to Catalogue 1. The pseudo redshift of each source was estimated via a Bayesian method to determine the isotropic-equivalent energy.

The $V_\text{max}$ method was adopted since the energy function and rate evolution can be derived without any prior model assumptions. After correcting volumetric rates with selection functions derived from BaseCat samples, the energy function within different redshift bins follows a Schechter function, with a slope index of $\gamma \in [-1.8, -1.6]$. However, the lack of low-energy samples from far distances leaves the high-redshift slope poorly constrained.

The evolution of the FRB volumetric event rate remains uncertain. We find an FRB event rate of $2.67\times10^{4}~\text{Gpc}^{-3}~\text{yr}^{-1}$ above $10^{38}~\text{erg}$ at $z=0.14$. The possibility that FRBs evolve with the SFRD or SMD scenario cannot be ruled out. Considering the possibility that FRBs are primarily produced from BNS mergers or magnetars from different formation channels, their population may evolve with a delay relative to the SFRD. Nevertheless, we tested the delay model and find that the FRB event rate evolution is possibly consistent with a delay model with a delay time of $\tau\sim4.66$ Gyr.

In summary, our analysis emphasises the importance of accurate fluence measurements and energy estimates for probing FRB properties and their event rate evolution. While our study provides valuable insights using CHIME/FRB baseband data, our results are limited by the BaseCat 1 sample size and baseband system. Our study highlights the importance of large-sample baseband data, multi-wavelength data, and comprehensive selection functions to better reveal FRB properties. The baseband data from CHIME/FRB Catalogue 2 \citep{2026ApJS..283...34C} will be essential for investigating FRB populations. Fast radio bursts with $z>2$ are expected to be particularly valuable for distinguishing progenitor scenarios and constraining volumetric rate evolution \citep{2025arXiv250801648C}. Multi-wavelength follow-up observations of well-localised FRBs will also provide crucial tests of progenitor models.

%%%%%%%%%%%%%%%%%%%%%%%%%%%%%%%%%%%%%%%%%%%%%%%%%%%%%%%%%%%%%%
\begin{acknowledgements}

This work was supported by the National Natural Science Foundation of China (grant No. 12288102), the Natural Science Foundation of Xinjiang Uygur Autonomous Region (2022D01D85), the Major Science and Technology Program of Xinjiang Uygur Autonomous Region (2022A03013-2), the Tianchi Talent project (Ju-Mei Yao), the Chinese Academy of Sciences (CAS) Project for Young Scientists in Basic Research (YSBR-063), the Tianshan talents program (2023TSYCTD0013), and the CAS “Light of West China” Program (Nos. xbzg-zdsys-202410 and 2022-XBQNXZ-015). The work was also supported by the National Natural Science Foundation of China (grant Nos. 12573052, 12573103, 12494575, 12273009, 12373114).

\end{acknowledgements}

% Please note that we have included the references below in
% order to compile the document, but we ask you to:

\bibliographystyle{bibtex/aa}
\bibliography{citations}

\begin{thebibliography}{86}
\expandafter\ifx\csname natexlab\endcsname\relax\def\natexlab#1{#1}\fi

\bibitem[{{Ai} {et~al.}(2021){Ai}, {Gao}, \& {Zhang}}]{2021ApJ...906L...5A}
{Ai}, S., {Gao}, H., \& {Zhang}, B. 2021, \apjl, 906, L5

\bibitem[{{Arcus} {et~al.}(2025){Arcus}, {James}, {Ekers}, {Macquart}, {Sadler}, {Wayth}, {Bannister}, {Deller}, {Flynn}, {Glowacki}, \& et~al.}]{2025PASA...42....3A}
{Arcus}, W.~R., {James}, C., {Ekers}, R., {et~al.} 2025, \pasa, 42, e003

\bibitem[{{Ashton} {et~al.}(2019){Ashton}, {H{\"u}bner}, {Lasky}, {Talbot}, {Ackley}, {Biscoveanu}, {Chu}, {Divakarla}, {Easter}, {Goncharov}, {Hernandez Vivanco}, {Harms}, {Lower}, {Meadors}, {Melchor}, {Payne}, {Pitkin}, {Powell}, {Sarin}, {Smith}, \& {Thrane}}]{2019ApJS..241...27A}
{Ashton}, G., {H{\"u}bner}, M., {Lasky}, P.~D., {et~al.} 2019, ApJS, 241, 27

\bibitem[{{Avni} \& {Bahcall}(1980)}]{1980ApJ...235..694A}
{Avni}, Y. \& {Bahcall}, J.~N. 1980, \apj, 235, 694

\bibitem[{{Bannister} {et~al.}(2017){Bannister}, {Shannon}, {Macquart}, {Flynn}, {Edwards}, {O'Neill}, {Os{\l}owski}, {Bailes}, {Zackay}, {Clarke}, {D'Addario}, {Dodson}, {Hall}, {Jameson}, {Jones}, {Navarro}, {Trinh}, {Allison}, {Anderson}, {Bell}, {Chippendale}, {Collier}, {Heald}, {Heywood}, {Hotan}, {Lee-Waddell}, {Madrid}, {Marvil}, {McConnell}, {Popping}, {Voronkov}, {Whiting}, {Allen}, {Bock}, {Brodrick}, {Cooray}, {DeBoer}, {Diamond}, {Ekers}, {Gough}, {Hampson}, {Harvey-Smith}, {Hay}, {Hayman}, {Jackson}, {Johnston}, {Koribalski}, {McClure-Griffiths}, {Mirtschin}, {Ng}, {Norris}, {Pearce}, {Phillips}, {Roxby}, {Troup}, \& {Westmeier}}]{2017ApJ...841L..12B}
{Bannister}, K.~W., {Shannon}, R.~M., {Macquart}, J.~P., {et~al.} 2017, \apjl, 841, L12

\bibitem[{{Bhandari} {et~al.}(2022){Bhandari}, {Heintz}, {Aggarwal}, {Marnoch}, {Day}, {Sydnor}, {Burke-Spolaor}, {Law}, {Xavier Prochaska}, {Tejos}, {Bannister}, {Butler}, {Deller}, {Ekers}, {Flynn}, {Fong}, {James}, {Lazio}, {Luo}, {Mahony}, {Ryder}, {Sadler}, {Shannon}, {Han}, {Lee}, \& {Zhang}}]{2022AJ....163...69B}
{Bhandari}, S., {Heintz}, K.~E., {Aggarwal}, K., {et~al.} 2022, \apj, 163, 69

\bibitem[{{Bhandari} {et~al.}(2020){Bhandari}, {Sadler}, {Prochaska}, {Simha}, {Ryder}, {Marnoch}, {Bannister}, {Macquart}, {Flynn}, {Shannon}, {Tejos}, {Corro-Guerra}, {Day}, {Deller}, {Ekers}, {Lopez}, {Mahony}, {Nu{\~n}ez}, \& {Phillips}}]{2020ApJ...895L..37B}
{Bhandari}, S., {Sadler}, E.~M., {Prochaska}, J.~X., {et~al.} 2020, \apjl, 895, L37

\bibitem[{{Bhardwaj} {et~al.}(2025){Bhardwaj}, {Kaspi}, {Masui}, {Gaensler}, {Ibik}, \& {Sammons}}]{2025ApJ...989..130B}
{Bhardwaj}, M., {Kaspi}, V.~M., {Masui}, K.~W., {et~al.} 2025, \apj, 989, 130

\bibitem[{{Bochenek} {et~al.}(2020){Bochenek}, {Ravi}, {Belov}, {Hallinan}, {Kocz}, {Kulkarni}, \& {McKenna}}]{2020Natur.587...59B}
{Bochenek}, C.~D., {Ravi}, V., {Belov}, K.~V., {et~al.} 2020, Nat, 587, 59

\bibitem[{{Caleb} {et~al.}(2025){Caleb}, {Nanayakkara}, {Stappers}, {Pastor-Marazuela}, {Khrykin}, {Glazebrook}, {Tejos}, {Prochaska}, {Rajwade}, {Mas-Ribas}, {Driessen}, {Fong}, {Gordon}, {Hoffmann}, {James}, {Jankowski}, {Kahinga}, {Kramer}, {Simha}, {Barr}, {Christiaan Bezuidenhout}, {Deng}, {Lin}, {Marnoch}, {Martin}, {Nugent}, {Shaji}, \& {Tian}}]{2025arXiv250801648C}
{Caleb}, M., {Nanayakkara}, T., {Stappers}, B., {et~al.} 2025, arXiv e-prints, arXiv:2508.01648

\bibitem[{{Chen} {et~al.}(2024){Chen}, {Jia}, {Dong}, \& {Wang}}]{2024ApJ...973L..54C}
{Chen}, J.~H., {Jia}, X.~D., {Dong}, X.~F., \& {Wang}, F.~Y. 2024, \apjl, 973, L54

\bibitem[{{Chennamangalam} {et~al.}(2013){Chennamangalam}, {Lorimer}, {Mandel}, \& {Bagchi}}]{2013MNRAS.431..874C}
{Chennamangalam}, J., {Lorimer}, D.~R., {Mandel}, I., \& {Bagchi}, M. 2013, \mnras, 431, 874

\bibitem[{{CHIME/FRB Collaboration} {et~al.}(2026){CHIME/FRB Collaboration}, {Abbott}, {Andersen}, {Andrew}, {Bandura}, {Bhardwaj}, {Bhusare}, {Brar}, {Cassanelli}, {Chatterjee}, {Cliche}, {Cook}, {Curtin}, {Dobbs}, {Dong}, {Eadie}, {Eftekhari}, {Fonseca}, {Gaensler}, {Good}, {Halpern}, {Hessels}, {Ibik}, {Jain}, {Joseph}, {Kader}, {Kaspi}, {Khan}, {Kharel}, {Kumar}, {Landecker}, {Lang}, {Lanman}, {L'Argent}, {Lazda}, {Leung}, {Li}, {Lintott}, {Main}, {Masui}, {Mate}, {McGregor}, {McKinven}, {Mena-Parra}, {Meyers}, {Michilli}, {Ng}, {Ng}, {Nimmo}, {Noble}, {Pandhi}, {Patil}, {Pearlman}, {Pen}, {Pleunis}, {Prochaska}, {Rafiei-Ravandi}, {Ransom}, {Renard}, {Sammons}, {Sand}, {Scholz}, {Shah}, {Shin}, {Siegel}, {Sirota}, {Smith}, {Stairs}, {Stenning}, {Tendulkar}, {Vanderlinde}, {Walmsley}, {Wang}, \& {Wulf}}]{2026ApJS..283...34C}
{CHIME/FRB Collaboration}, {Abbott}, T., {Andersen}, B.~C., {et~al.} 2026, \apjs, 283, 34

\bibitem[{{CHIME/FRB Collaboration} {et~al.}(2024){CHIME/FRB Collaboration}, {Amiri}, {Andersen}, {Andrew}, {Bandura}, {Bhardwaj}, {Boyle}, {Brar}, {Breitman}, {Cassanelli}, {Chawla}, {Cook}, {Curtin}, {Dobbs}, {Dong}, {Eadie}, {Fonseca}, {Gaensler}, {Giri}, {Herrera-Martin}, {Hopkins}, {Ibik}, {Joseph}, {Kaczmarek}, {Kader}, {Kaspi}, {Lanman}, {Lazda}, {Leung}, {Liu}, {Masui}, {McKinven}, {Mena-Parra}, {Merryfield}, {Michilli}, {Ng}, {Nimmo}, {Noble}, {Pandhi}, {Patel}, {Pearlman}, {Pen}, {Petroff}, {Pleunis}, {Rafiei-Ravandi}, {Rahman}, {Ransom}, {Sand}, {Scholz}, {Shah}, {Shin}, {Shpunarska}, {Siegel}, {Smith}, {Stairs}, {Stenning}, {Vanderlinde}, {Wang}, {White}, \& {Wulf}}]{2024ApJ...969..145C}
{CHIME/FRB Collaboration}, {Amiri}, M., {Andersen}, B.~C., {et~al.} 2024, \apj, 969, 145

\bibitem[{{CHIME/FRB Collaboration} {et~al.}(2021){CHIME/FRB Collaboration}, {Amiri}, {Andersen}, {Bandura}, {Berger}, {Bhardwaj}, {Boyce}, {Boyle}, {Brar}, {Breitman}, {Cassanelli}, {Chawla}, {Chen}, {Cliche}, {Cook}, {Cubranic}, {Curtin}, {Deng}, {Dobbs}, {Dong}, {Eadie}, {Fandino}, {Fonseca}, {Gaensler}, {Giri}, {Good}, {Halpern}, {Hill}, {Hinshaw}, {Josephy}, {Kaczmarek}, {Kader}, {Kania}, {Kaspi}, {Landecker}, {Lang}, {Leung}, {Li}, {Lin}, {Masui}, {McKinven}, {Mena-Parra}, {Merryfield}, {Meyers}, {Michilli}, {Milutinovic}, {Mirhosseini}, {M{\"u}nchmeyer}, {Naidu}, {Newburgh}, {Ng}, {Patel}, {Pen}, {Petroff}, {Pinsonneault-Marotte}, {Pleunis}, {Rafiei-Ravandi}, {Rahman}, {Ransom}, {Renard}, {Sanghavi}, {Scholz}, {Shaw}, {Shin}, {Siegel}, {Sikora}, {Singh}, {Smith}, {Stairs}, {Tan}, {Tendulkar}, {Vanderlinde}, {Wang}, {Wulf}, \& {Zwaniga}}]{2021ApJS..257...59C}
{CHIME/FRB Collaboration}, {Amiri}, M., {Andersen}, B.~C., {et~al.} 2021, \apjs, 257, 59

\bibitem[{{CHIME/FRB Collaboration} {et~al.}(2018){CHIME/FRB Collaboration}, {Amiri}, {Bandura}, {Berger}, {Bhardwaj}, {Boyce}, {Boyle}, {Brar}, {Burhanpurkar}, {Chawla}, {Chowdhury}, {Cliche}, {Cranmer}, {Cubranic}, {Deng}, {Denman}, {Dobbs}, {Fandino}, {Fonseca}, {Gaensler}, {Giri}, {Gilbert}, {Good}, {Guliani}, {Halpern}, {Hinshaw}, {H{\"o}fer}, {Josephy}, {Kaspi}, {Landecker}, {Lang}, {Liao}, {Masui}, {Mena-Parra}, {Naidu}, {Newburgh}, {Ng}, {Patel}, {Pen}, {Pinsonneault-Marotte}, {Pleunis}, {Rafiei Ravandi}, {Ransom}, {Renard}, {Scholz}, {Sigurdson}, {Siegel}, {Smith}, {Stairs}, {Tendulkar}, {Vanderlinde}, \& {Wiebe}}]{2018ApJ...863...48C}
{CHIME/FRB Collaboration}, {Amiri}, M., {Bandura}, K., {et~al.} 2018, \apj, 863, 48

\bibitem[{{CHIME/FRB Collaboration} {et~al.}(2019{\natexlab{a}}){CHIME/FRB Collaboration}, {Amiri}, {Bandura}, {Bhardwaj}, {Boubel}, {Boyce}, {Boyle}, {. Brar}, {Burhanpurkar}, {Cassanelli}, {Chawla}, {Cliche}, {Cubranic}, {Deng}, {Denman}, {Dobbs}, {Fandino}, {Fonseca}, {Gaensler}, {Gilbert}, {Gill}, {Giri}, {Good}, {Halpern}, {Hanna}, {Hill}, {Hinshaw}, {H{\"o}fer}, {Josephy}, {Kaspi}, {Landecker}, {Lang}, {Lin}, {Masui}, {Mckinven}, {Mena-Parra}, {Merryfield}, {Michilli}, {Milutinovic}, {Moatti}, {Naidu}, {Newburgh}, {Ng}, {Patel}, {Pen}, {Pinsonneault-Marotte}, {Pleunis}, {Rafiei-Ravandi}, {Rahman}, {Ransom}, {Renard}, {Scholz}, {Shaw}, {Siegel}, {Smith}, {Stairs}, {Tendulkar}, {Tretyakov}, {Vanderlinde}, \& {Yadav}}]{2019Natur.566..235C}
{CHIME/FRB Collaboration}, {Amiri}, M., {Bandura}, K., {et~al.} 2019{\natexlab{a}}, Nat, 566, 235

\bibitem[{{CHIME/FRB Collaboration} {et~al.}(2019{\natexlab{b}}){CHIME/FRB Collaboration}, {Amiri}, {Bandura}, {Bhardwaj}, {Boubel}, {Boyce}, {Boyle}, {Brar}, {Burhanpurkar}, {Chawla}, {Cliche}, {Cubranic}, {Deng}, {Denman}, {Dobbs}, {Fandino}, {Fonseca}, {Gaensler}, {Gilbert}, {Giri}, {Good}, {Halpern}, {Hanna}, {Hill}, {Hinshaw}, {H{\"o}fer}, {Josephy}, {Kaspi}, {Landecker}, {Lang}, {Masui}, {Mckinven}, {Mena-Parra}, {Merryfield}, {Milutinovic}, {Moatti}, {Naidu}, {Newburgh}, {Ng}, {Patel}, {Pen}, {Pinsonneault-Marotte}, {Pleunis}, {Rafiei-Ravandi}, {Ransom}, {Renard}, {Scholz}, {Shaw}, {Siegel}, {Smith}, {Stairs}, {Tendulkar}, {Tretyakov}, {Vanderlinde}, \& {Yadav}}]{2019Natur.566..230C}
{CHIME/FRB Collaboration}, {Amiri}, M., {Bandura}, K., {et~al.} 2019{\natexlab{b}}, Nat, 566, 230

\bibitem[{{CHIME/FRB Collaboration} {et~al.}(2019{\natexlab{c}}){CHIME/FRB Collaboration}, {Andersen}, {Bandura}, {Bhardwaj}, {Boubel}, {Boyce}, {Boyle}, {Brar}, {Cassanelli}, {Chawla}, {Cubranic}, {Deng}, {Dobbs}, {Fandino}, {Fonseca}, {Gaensler}, {Gilbert}, {Giri}, {Good}, {Halpern}, {Hill}, {Hinshaw}, {H{\"o}fer}, {Josephy}, {Kaspi}, {Kothes}, {Landecker}, {Lang}, {Li}, {Lin}, {Masui}, {Mena-Parra}, {Merryfield}, {Mckinven}, {Michilli}, {Milutinovic}, {Naidu}, {Newburgh}, {Ng}, {Patel}, {Pen}, {Pinsonneault-Marotte}, {Pleunis}, {Rafiei-Ravandi}, {Rahman}, {Ransom}, {Renard}, {Scholz}, {Siegel}, {Singh}, {Smith}, {Stairs}, {Tendulkar}, {Tretyakov}, {Vanderlinde}, {Yadav}, \& {Zwaniga}}]{2019ApJ...885L..24C}
{CHIME/FRB Collaboration}, {Andersen}, B.~C., {Bandura}, K., {et~al.} 2019{\natexlab{c}}, \apjl, 885, L24

\bibitem[{{CHIME/FRB Collaboration} {et~al.}(2023){CHIME/FRB Collaboration}, {Andersen}, {Bandura}, {Bhardwaj}, {Boyle}, {Brar}, {Cassanelli}, {Chatterjee}, {Chawla}, {Cook}, {Curtin}, {Dobbs}, {Dong}, {Faber}, {Fandino}, {Fonseca}, {Gaensler}, {Giri}, {Herrera-Martin}, {Hill}, {Ibik}, {Josephy}, {Kaczmarek}, {Kader}, {Kaspi}, {Landecker}, {Lanman}, {Lazda}, {Leung}, {Lin}, {Masui}, {McKinven}, {Mena-Parra}, {Meyers}, {Michilli}, {Ng}, {Pandhi}, {Pearlman}, {Pen}, {Petroff}, {Pleunis}, {Rafiei-Ravandi}, {Rahman}, {Ransom}, {Renard}, {Sand}, {Sanghavi}, {Scholz}, {Shah}, {Shin}, {Siegel}, {Smith}, {Stairs}, {Su}, {Tendulkar}, {Vanderlinde}, {Wang}, {Wulf}, \& {Zwaniga}}]{2023ApJ...947...83C}
{CHIME/FRB Collaboration}, {Andersen}, B.~C., {Bandura}, K., {et~al.} 2023, \apj, 947, 83

\bibitem[{{CHIME/FRB Collaboration} {et~al.}(2020){CHIME/FRB Collaboration}, {Andersen}, {Bandura}, {Bhardwaj}, {Bij}, {Boyce}, {Boyle}, {Brar}, {Cassanelli}, {Chawla}, {Chen}, {Cliche}, {Cook}, {Cubranic}, {Curtin}, {Denman}, {Dobbs}, {Dong}, {Fandino}, {Fonseca}, {Gaensler}, {Giri}, {Good}, {Halpern}, {Hill}, {Hinshaw}, {H{\"o}fer}, {Josephy}, {Kania}, {Kaspi}, {Landecker}, {Leung}, {Li}, {Lin}, {Masui}, {McKinven}, {Mena-Parra}, {Merryfield}, {Meyers}, {Michilli}, {Milutinovic}, {Mirhosseini}, {M{\"u}nchmeyer}, {Naidu}, {Newburgh}, {Ng}, {Patel}, {Pen}, {Pinsonneault-Marotte}, {Pleunis}, {Quine}, {Rafiei-Ravandi}, {Rahman}, {Ransom}, {Renard}, {Sanghavi}, {Scholz}, {Shaw}, {Shin}, {Siegel}, {Singh}, {Smegal}, {Smith}, {Stairs}, {Tan}, {Tendulkar}, {Tretyakov}, {Vanderlinde}, {Wang}, {Wulf}, \& {Zwaniga}}]{2020Natur.587...54C}
{CHIME/FRB Collaboration}, {Andersen}, B.~C., {Bandura}, K.~M., {et~al.} 2020, Nat, 587, 54

\bibitem[{{Cordes} \& {McLaughlin}(2003)}]{2003ApJ...596.1142C}
{Cordes}, J.~M. \& {McLaughlin}, M.~A. 2003, \apj, 596, 1142

\bibitem[{{Cui} {et~al.}(2025){Cui}, {James}, {Li}, \& {Zhang}}]{2025ApJ...982..158C}
{Cui}, X.-H., {James}, C.~W., {Li}, D., \& {Zhang}, C.-M. 2025, \apj, 982, 158

\bibitem[{{Cui} {et~al.}(2022){Cui}, {Zhang}, {Li}, {Zhang}, {Peng}, {Zhu}, {Strom}, {Wang}, {Wang}, {Wu}, \& et~al.}]{2022Ap&SS.367...66C}
{Cui}, X.-H., {Zhang}, C.-M., {Li}, D., {et~al.} 2022, \apss, 367, 66

\bibitem[{{Cui} {et~al.}(2021){Cui}, {Zhang}, {Wang}, {Zhang}, {Li}, {Peng}, {Zhu}, {Wang}, {Strom}, {Ye}, {Wang}, \& {Yang}}]{2021MNRAS.500.3275C}
{Cui}, X.-H., {Zhang}, C.-M., {Wang}, S.-Q., {et~al.} 2021, \mnras, 500, 3275

\bibitem[{{Foreman-Mackey} {et~al.}(2013){Foreman-Mackey}, {Hogg}, {Lang}, \& {Goodman}}]{2013PASP..125..306F}
{Foreman-Mackey}, D., {Hogg}, D.~W., {Lang}, D., \& {Goodman}, J. 2013, \pasp, 125, 306

\bibitem[{{Gajjar} {et~al.}(2018){Gajjar}, {Siemion}, {Price}, {Law}, {Michilli}, {Hessels}, {Chatterjee}, {Archibald}, {Bower}, {Brinkman}, {Burke-Spolaor}, {Cordes}, {Croft}, {Enriquez}, {Foster}, {Gizani}, {Hellbourg}, {Isaacson}, {Kaspi}, {Lazio}, {Lebofsky}, {Lynch}, {MacMahon}, {McLaughlin}, {Ransom}, {Scholz}, {Seymour}, {Spitler}, {Tendulkar}, {Werthimer}, \& {Zhang}}]{2018ApJ...863....2G}
{Gajjar}, V., {Siemion}, A.~P.~V., {Price}, D.~C., {et~al.} 2018, \apj, 863, 2

\bibitem[{{Gupta} {et~al.}(2025){Gupta}, {Beniamini}, {Kumar}, \& {Finkelstein}}]{2025ApJ...986..100G}
{Gupta}, O., {Beniamini}, P., {Kumar}, P., \& {Finkelstein}, S.~L. 2025, \apj, 986, 100

\bibitem[{{Hallinan} {et~al.}(2019){Hallinan}, {Ravi}, {Weinreb}, {Kocz}, {Huang}, {Woody}, {Lamb}, {D'Addario}, {Catha}, {Law}, \& et~al.}]{2019BAAS...51g.255H}
{Hallinan}, G., {Ravi}, V., {Weinreb}, S., {et~al.} 2019, in Bulletin of the American Astronomical Society, Vol.~51, 255

\bibitem[{{Hashimoto} {et~al.}(2022){Hashimoto}, {Goto}, {Chen}, {Ho}, {Hsiao}, {Wong}, {On}, {Kim}, {Kilerci-Eser}, {Huang}, {Santos}, \& {Yamasaki}}]{2022MNRAS.511.1961H}
{Hashimoto}, T., {Goto}, T., {Chen}, B.~H., {et~al.} 2022, MNRAS, 511, 1961

\bibitem[{{Hashimoto} {et~al.}(2020{\natexlab{a}}){Hashimoto}, {Goto}, {On}, {Lu}, {Santos}, {Ho}, {Kim}, {Wang}, \& {Hsiao}}]{2020MNRAS.498.3927H}
{Hashimoto}, T., {Goto}, T., {On}, A. Y.~L., {et~al.} 2020{\natexlab{a}}, MNRAS, 498, 3927

\bibitem[{{Hashimoto} {et~al.}(2020{\natexlab{b}}){Hashimoto}, {Goto}, {Wang}, {Kim}, {Ho}, {On}, {Lu}, \& {Santos}}]{2020MNRAS.494.2886H}
{Hashimoto}, T., {Goto}, T., {Wang}, T.-W., {et~al.} 2020{\natexlab{b}}, \mnras, 494, 2886

\bibitem[{{Higson} {et~al.}(2019){Higson}, {Handley}, {Hobson}, \& {Lasenby}}]{2019S&C....29..891H}
{Higson}, E., {Handley}, W., {Hobson}, M., \& {Lasenby}, A. 2019, S\&C, 29, 891

\bibitem[{{Hoffmann} {et~al.}(2025){Hoffmann}, {James}, {Glowacki}, {Prochaska}, {Gordon}, {Deller}, {Shannon}, \& {Ryder}}]{2025PASA...42...17H}
{Hoffmann}, J.~L., {James}, C., {Glowacki}, M., {et~al.} 2025, \pasa, 42, e017

\bibitem[{{Ibik} {et~al.}(2024){Ibik}, {Drout}, {Gaensler}, {Scholz}, {Michilli}, {Bhardwaj}, {Kaspi}, {Pleunis}, {Cassanelli}, {Cook}, \& et~al.}]{2024ApJ...961...99I}
{Ibik}, A.~L., {Drout}, M.~R., {Gaensler}, B.~M., {et~al.} 2024, \apj, 961, 99

\bibitem[{{Ivezi{\'c}} {et~al.}(2020){Ivezi{\'c}}, {Connolly}, {VanderPlas}, \& {Gray}}]{2020sdmm.book.....I}
{Ivezi{\'c}}, {\v{Z}}., {Connolly}, A.~J., {VanderPlas}, J.~T., \& {Gray}, A. 2020, {Statistics, Data Mining, and Machine Learning in Astronomy. A Practical Python Guide for the Analysis of Survey Data, Updated Edition}

\bibitem[{{James} {et~al.}(2022{\natexlab{a}}){James}, {Prochaska}, {Macquart}, {North-Hickey}, {Bannister}, \& {Dunning}}]{2022MNRAS.510L..18J}
{James}, C.~W., {Prochaska}, J.~X., {Macquart}, J.~P., {et~al.} 2022{\natexlab{a}}, MNRAS, 510, L18

\bibitem[{{James} {et~al.}(2022{\natexlab{b}}){James}, {Prochaska}, {Macquart}, {North-Hickey}, {Bannister}, \& {Dunning}}]{2022MNRAS.509.4775J}
{James}, C.~W., {Prochaska}, J.~X., {Macquart}, J.~P., {et~al.} 2022{\natexlab{b}}, \mnras, 509, 4775

\bibitem[{{Kirsten} {et~al.}(2024){Kirsten}, {Ould-Boukattine}, {Herrmann}, {Gawro{\'n}ski}, {Hessels}, {Lu}, {Snelders}, {Chawla}, {Yang}, {Blaauw}, \& et~al.}]{2024NatAs...8..337K}
{Kirsten}, F., {Ould-Boukattine}, O.~S., {Herrmann}, W., {et~al.} 2024, NatAs, 8, 337

\bibitem[{{Li} {et~al.}(2021){Li}, {Lin}, {Xiong}, {Ge}, {Li}, {Li}, {Lu}, {Zhang}, {Tuo}, {Nang}, {Zhang}, {Xiao}, {Chen}, {Song}, {Xu}, {Liu}, {Jia}, {Cao}, {Qu}, {Zhang}, {Gu}, {Liao}, {Zhao}, {Tan}, {Nie}, {Zhao}, {Zheng}, {Zheng}, {Luo}, {Cai}, {Li}, {Xue}, {Bu}, {Chang}, {Chen}, {Chen}, {Chen}, {Chen}, {Chen}, {Cui}, {Cui}, {Deng}, {Dong}, {Du}, {Fu}, {Gao}, {Gao}, {Gao}, {Gu}, {Guan}, {Guo}, {Han}, {Huang}, {Huo}, {Jiang}, {Jiang}, {Jin}, {Jin}, {Kong}, {Li}, {Li}, {Li}, {Li}, {Li}, {Li}, {Li}, {Liang}, {Liu}, {Liu}, {Liu}, {Liu}, {Liu}, {Lu}, {Lu}, {Luo}, {Ma}, {Meng}, {Ou}, {Sai}, {Shang}, {Song}, {Sun}, {Tao}, {Wang}, {Wang}, {Wang}, {Wang}, {Wang}, {Wen}, {Wu}, {Wu}, {Wu}, {Xiao}, {Xu}, {Yang}, {Yang}, {Yang}, {Yang}, {Yi}, {Yin}, {You}, {Zhang}, {Zhang}, {Zhang}, {Zhang}, {Zhang}, {Zhang}, {Zhang}, {Zhang}, {Zhang}, {Zhang}, {Zhang}, {Zhang}, {Zhang}, {Zhang}, {Zhang}, {Zhang}, {Zhou}, {Zhou}, {Zhu}, {Zhu}, \& {Zhuang}}]{2021NatAs...5..378L}
{Li}, C.~K., {Lin}, L., {Xiong}, S.~L., {et~al.} 2021, NatAs, 5, 378

\bibitem[{{L{\'o}pez Fern{\'a}ndez} {et~al.}(2018){L{\'o}pez Fern{\'a}ndez}, {Gonz{\'a}lez Delgado}, {P{\'e}rez}, {Garc{\'\i}a-Benito}, {Cid Fernandes}, {Schoenell}, {S{\'a}nchez}, {Gallazzi}, {S{\'a}nchez-Bl{\'a}zquez}, {Vale Asari}, \& {Walcher}}]{2018A&A...615A..27L}
{L{\'o}pez Fern{\'a}ndez}, R., {Gonz{\'a}lez Delgado}, R.~M., {P{\'e}rez}, E., {et~al.} 2018, \aap, 615, A27

\bibitem[{{Lorimer} {et~al.}(2007){Lorimer}, {Bailes}, {McLaughlin}, {Narkevic}, \& {Crawford}}]{2007Sci...318..777L}
{Lorimer}, D.~R., {Bailes}, M., {McLaughlin}, M.~A., {Narkevic}, D.~J., \& {Crawford}, F. 2007, Sci, 318, 777

\bibitem[{{Luo} {et~al.}(2020){Luo}, {Men}, {Lee}, {Wang}, {Lorimer}, \& {Zhang}}]{2020MNRAS.494..665L}
{Luo}, R., {Men}, Y., {Lee}, K., {et~al.} 2020, MNRAS, 494, 665

\bibitem[{{Lynden-Bell}(1971)}]{1971MNRAS.155...95L}
{Lynden-Bell}, D. 1971, \mnras, 155, 95

\bibitem[{{Ma} {et~al.}(2025){Ma}, {Gao}, {Li}, {Niu}, {Yao}, \& {Wang}}]{2025ApJ...981...24M}
{Ma}, W.~Q., {Gao}, Z.~F., {Li}, B.~P., {et~al.} 2025, \apj, 981, 24

\bibitem[{{Macquart} \& {Ekers}(2018{\natexlab{a}})}]{2018MNRAS.480.4211M}
{Macquart}, J.~P. \& {Ekers}, R. 2018{\natexlab{a}}, \mnras, 480, 4211

\bibitem[{{Macquart} \& {Ekers}(2018{\natexlab{b}})}]{2018MNRAS.474.1900M}
{Macquart}, J.-P. \& {Ekers}, R.~D. 2018{\natexlab{b}}, \mnras, 474, 1900

\bibitem[{{Macquart} {et~al.}(2020){Macquart}, {Prochaska}, {McQuinn}, {Bannister}, {Bhandari}, {Day}, {Deller}, {Ekers}, {James}, {Marnoch}, {Os{\l}owski}, {Phillips}, {Ryder}, {Scott}, {Shannon}, \& {Tejos}}]{2020Natur.581..391M}
{Macquart}, J.~P., {Prochaska}, J.~X., {McQuinn}, M., {et~al.} 2020, Nat, 581, 391

\bibitem[{{Macquart} {et~al.}(2019){Macquart}, {Shannon}, {Bannister}, {James}, {Ekers}, \& {Bunton}}]{2019ApJ...872L..19M}
{Macquart}, J.~P., {Shannon}, R.~M., {Bannister}, K.~W., {et~al.} 2019, \apjl, 872, L19

\bibitem[{{Madau} \& {Dickinson}(2014)}]{2014ARA&A..52..415M}
{Madau}, P. \& {Dickinson}, M. 2014, \araa, 52, 415

\bibitem[{{Madau} \& {Fragos}(2017)}]{2017ApJ...840...39M}
{Madau}, P. \& {Fragos}, T. 2017, \apj, 840, 39

\bibitem[{{Mereghetti} {et~al.}(2020){Mereghetti}, {Savchenko}, {Ferrigno}, {G{\"o}tz}, {Rigoselli}, {Tiengo}, {Bazzano}, {Bozzo}, {Coleiro}, {Courvoisier}, {Doyle}, {Goldwurm}, {Hanlon}, {Jourdain}, {von Kienlin}, {Lutovinov}, {Martin-Carrillo}, {Molkov}, {Natalucci}, {Onori}, {Panessa}, {Rodi}, {Rodriguez}, {S{\'a}nchez-Fern{\'a}ndez}, {Sunyaev}, \& {Ubertini}}]{2020ApJ...898L..29M}
{Mereghetti}, S., {Savchenko}, V., {Ferrigno}, C., {et~al.} 2020, \apjl, 898, L29

\bibitem[{{Merryfield} {et~al.}(2023){Merryfield}, {Tendulkar}, {Shin}, {Andersen}, {Josephy}, {Good}, {Dong}, {Masui}, {Lang}, {M{\"u}nchmeyer}, \& et~al.}]{2023AJ....165..152M}
{Merryfield}, M., {Tendulkar}, S.~P., {Shin}, K., {et~al.} 2023, \aj, 165, 152

\bibitem[{{Mo} {et~al.}(2023){Mo}, {Zhu}, {Wang}, {Tang}, \& {Feng}}]{2023MNRAS.518..539M}
{Mo}, J.-F., {Zhu}, W., {Wang}, Y., {Tang}, L., \& {Feng}, L.-L. 2023, \mnras, 518, 539

\bibitem[{{Moroianu} {et~al.}(2023){Moroianu}, {Wen}, {James}, {Ai}, {Kovalam}, {Panther}, \& {Zhang}}]{2023NatAs...7..579M}
{Moroianu}, A., {Wen}, L., {James}, C.~W., {et~al.} 2023, NatAs, 7, 579

\bibitem[{{Pastor-Marazuela} {et~al.}(2021){Pastor-Marazuela}, {Connor}, {van Leeuwen}, {Maan}, {ter Veen}, {Bilous}, {Oostrum}, {Petroff}, {Straal}, {Vohl}, {Attema}, {Boersma}, {Kooistra}, {van der Schuur}, {Sclocco}, {Smits}, {Adams}, {Adebahr}, {de Blok}, {Coolen}, {Damstra}, {D{\'e}nes}, {Hess}, {van der Hulst}, {Hut}, {Ivashina}, {Kutkin}, {Loose}, {Lucero}, {Mika}, {Moss}, {Mulder}, {Norden}, {Oosterloo}, {Orr{\'u}}, {Ruiter}, \& {Wijnholds}}]{2021Natur.596..505P}
{Pastor-Marazuela}, I., {Connor}, L., {van Leeuwen}, J., {et~al.} 2021, Nat, 596, 505

\bibitem[{{Planck Collaboration} {et~al.}(2020){Planck Collaboration}, {Aghanim}, {Akrami}, {Ashdown}, {Aumont}, {Baccigalupi}, {Ballardini}, {Banday}, {Barreiro}, {Bartolo}, {Basak}, {Battye}, {Benabed}, {Bernard}, {Bersanelli}, {Bielewicz}, {Bock}, {Bond}, {Borrill}, {Bouchet}, {Boulanger}, {Bucher}, {Burigana}, {Butler}, {Calabrese}, {Cardoso}, {Carron}, {Challinor}, {Chiang}, {Chluba}, {Colombo}, {Combet}, {Contreras}, {Crill}, {Cuttaia}, {de Bernardis}, {de Zotti}, {Delabrouille}, {Delouis}, {Di Valentino}, {Diego}, {Dor{\'e}}, {Douspis}, {Ducout}, {Dupac}, {Dusini}, {Efstathiou}, {Elsner}, {En{\ss}lin}, {Eriksen}, {Fantaye}, {Farhang}, {Fergusson}, {Fernandez-Cobos}, {Finelli}, {Forastieri}, {Frailis}, {Fraisse}, {Franceschi}, {Frolov}, {Galeotta}, {Galli}, {Ganga}, {G{\'e}nova-Santos}, {Gerbino}, {Ghosh}, {Gonz{\'a}lez-Nuevo}, {G{\'o}rski}, {Gratton}, {Gruppuso}, {Gudmundsson}, {Hamann}, {Handley}, {Hansen}, {Herranz}, {Hildebrandt}, {Hivon}, {Huang}, {Jaffe}, {Jones}, {Karakci}, {Keih{\"a}nen},
  {Keskitalo}, {Kiiveri}, {Kim}, {Kisner}, {Knox}, {Krachmalnicoff}, {Kunz}, {Kurki-Suonio}, {Lagache}, {Lamarre}, {Lasenby}, {Lattanzi}, {Lawrence}, {Le Jeune}, {Lemos}, {Lesgourgues}, {Levrier}, {Lewis}, {Liguori}, {Lilje}, {Lilley}, {Lindholm}, {L{\'o}pez-Caniego}, {Lubin}, {Ma}, {Mac{\'\i}as-P{\'e}rez}, {Maggio}, {Maino}, {Mandolesi}, {Mangilli}, {Marcos-Caballero}, {Maris}, {Martin}, {Martinelli}, {Mart{\'\i}nez-Gonz{\'a}lez}, {Matarrese}, {Mauri}, {McEwen}, {Meinhold}, {Melchiorri}, {Mennella}, {Migliaccio}, {Millea}, {Mitra}, {Miville-Desch{\^e}nes}, {Molinari}, {Montier}, {Morgante}, {Moss}, {Natoli}, {N{\o}rgaard-Nielsen}, {Pagano}, {Paoletti}, {Partridge}, {Patanchon}, {Peiris}, {Perrotta}, {Pettorino}, {Piacentini}, {Polastri}, {Polenta}, {Puget}, {Rachen}, {Reinecke}, {Remazeilles}, {Renzi}, {Rocha}, {Rosset}, {Roudier}, {Rubi{\~n}o-Mart{\'\i}n}, {Ruiz-Granados}, {Salvati}, {Sandri}, {Savelainen}, {Scott}, {Shellard}, {Sirignano}, {Sirri}, {Spencer}, {Sunyaev}, {Suur-Uski}, {Tauber}, {Tavagnacco},
  {Tenti}, {Toffolatti}, {Tomasi}, {Trombetti}, {Valenziano}, {Valiviita}, {Van Tent}, {Vibert}, {Vielva}, {Villa}, {Vittorio}, {Wandelt}, {Wehus}, {White}, {White}, {Zacchei}, \& {Zonca}}]{2020A&A...641A...6P}
{Planck Collaboration}, {Aghanim}, N., {Akrami}, Y., {et~al.} 2020, \aap, 641, A6

\bibitem[{{Pleunis} {et~al.}(2021{\natexlab{a}}){Pleunis}, {Good}, {Kaspi}, {Mckinven}, {Ransom}, {Scholz}, {Bandura}, {Bhardwaj}, {Boyle}, {Brar}, {Cassanelli}, {Chawla}, {(Adam) Dong}, {Fonseca}, {Gaensler}, {Josephy}, {Kaczmarek}, {Leung}, {Lin}, {Masui}, {Mena-Parra}, {Michilli}, {Ng}, {Patel}, {Rafiei-Ravandi}, {Rahman}, {Sanghavi}, {Shin}, {Smith}, {Stairs}, \& {Tendulkar}}]{2021ApJ...923....1P}
{Pleunis}, Z., {Good}, D.~C., {Kaspi}, V.~M., {et~al.} 2021{\natexlab{a}}, \apj, 923, 1

\bibitem[{{Pleunis} {et~al.}(2021{\natexlab{b}}){Pleunis}, {Michilli}, {Bassa}, {Hessels}, {Naidu}, {Andersen}, {Chawla}, {Fonseca}, {Gopinath}, {Kaspi}, {Kondratiev}, {Li}, {Bhardwaj}, {Boyle}, {Brar}, {Cassanelli}, {Gupta}, {Josephy}, {Karuppusamy}, {Keimpema}, {Kirsten}, {Leung}, {Marcote}, {Masui}, {Mckinven}, {Meyers}, {Ng}, {Nimmo}, {Paragi}, {Rahman}, {Scholz}, {Shin}, {Smith}, {Stairs}, \& {Tendulkar}}]{2021ApJ...911L...3P}
{Pleunis}, Z., {Michilli}, D., {Bassa}, C.~G., {et~al.} 2021{\natexlab{b}}, \apjl, 911, L3

\bibitem[{{Ravi}(2019)}]{2019NatAs...3..928R}
{Ravi}, V. 2019, NatAs, 3, 928

\bibitem[{{Ridnaia} {et~al.}(2021){Ridnaia}, {Svinkin}, {Frederiks}, {Bykov}, {Popov}, {Aptekar}, {Golenetskii}, {Lysenko}, {Tsvetkova}, {Ulanov}, \& {Cline}}]{2021NatAs...5..372R}
{Ridnaia}, A., {Svinkin}, D., {Frederiks}, D., {et~al.} 2021, NatAs, 5, 372

\bibitem[{{Sand} {et~al.}(2025){Sand}, {Curtin}, {Michilli}, {Kaspi}, {Fonseca}, {Nimmo}, {Pleunis}, {Shin}, {Bhardwaj}, {Brar}, {Dobbs}, {Eadie}, {Gaensler}, {Joseph}, {Leung}, {Main}, {Masui}, {Mckinven}, {Pandhi}, {Pearlman}, {Rafiei-Ravandi}, {Sammons}, {Smith}, \& {Stairs}}]{2025ApJ...979..160S}
{Sand}, K.~R., {Curtin}, A.~P., {Michilli}, D., {et~al.} 2025, \apj, 979, 160

\bibitem[{{Schmidt}(1968)}]{1968ApJ...151..393S}
{Schmidt}, M. 1968, \apj, 151, 393

\bibitem[{{Shannon} {et~al.}(2018){Shannon}, {Macquart}, {Bannister}, {Ekers}, {James}, {Os{\l}owski}, {Qiu}, {Sammons}, {Hotan}, {Voronkov}, {Beresford}, {Brothers}, {Brown}, {Bunton}, {Chippendale}, {Haskins}, {Leach}, {Marquarding}, {McConnell}, {Pilawa}, {Sadler}, {Troup}, {Tuthill}, {Whiting}, {Allison}, {Anderson}, {Bell}, {Collier}, {G{\"u}rkan}, {Heald}, \& {Riseley}}]{2018Natur.562..386S}
{Shannon}, R.~M., {Macquart}, J.~P., {Bannister}, K.~W., {et~al.} 2018, Nat, 562, 386

\bibitem[{{Sharma} {et~al.}(2024){Sharma}, {Ravi}, {Connor}, {Law}, {Ocker}, {Sherman}, {Kosogorov}, {Faber}, {Hallinan}, {Harnach}, \& et~al.}]{2024Natur.635...61S}
{Sharma}, K., {Ravi}, V., {Connor}, L., {et~al.} 2024, Nat, 635, 61

\bibitem[{{Shin} {et~al.}(2023){Shin}, {Masui}, {Bhardwaj}, {Cassanelli}, {Chawla}, {Dobbs}, {Dong}, {Fonseca}, {Gaensler}, {Herrera-Mart{\'\i}n}, {Kaczmarek}, {Kaspi}, {Leung}, {Merryfield}, {Michilli}, {M{\"u}nchmeyer}, {Pearlman}, {Rafiei-Ravandi}, {Smith}, {Stairs}, \& {Tendulkar}}]{2023ApJ...944..105S}
{Shin}, K., {Masui}, K.~W., {Bhardwaj}, M., {et~al.} 2023, \apj, 944, 105

\bibitem[{{Shull} {et~al.}(2012){Shull}, {Smith}, \& {Danforth}}]{2012ApJ...759...23S}
{Shull}, J.~M., {Smith}, B.~D., \& {Danforth}, C.~W. 2012, \apj, 759, 23

\bibitem[{{Spitler} {et~al.}(2014){Spitler}, {Cordes}, {Hessels}, {Lorimer}, {McLaughlin}, {Chatterjee}, {Crawford}, {Deneva}, {Kaspi}, {Wharton}, {Allen}, {Bogdanov}, {Brazier}, {Camilo}, {Freire}, {Jenet}, {Karako-Argaman}, {Knispel}, {Lazarus}, {Lee}, {van Leeuwen}, {Lynch}, {Ransom}, {Scholz}, {Siemens}, {Stairs}, {Stovall}, {Swiggum}, {Venkataraman}, {Zhu}, {Aulbert}, \& {Fehrmann}}]{2014ApJ...790..101S}
{Spitler}, L.~G., {Cordes}, J.~M., {Hessels}, J.~W.~T., {et~al.} 2014, \apj, 790, 101

\bibitem[{{Spitler} {et~al.}(2016){Spitler}, {Scholz}, {Hessels}, {Bogdanov}, {Brazier}, {Camilo}, {Chatterjee}, {Cordes}, {Crawford}, {Deneva}, {Ferdman}, {Freire}, {Kaspi}, {Lazarus}, {Lynch}, {Madsen}, {McLaughlin}, {Patel}, {Ransom}, {Seymour}, {Stairs}, {Stappers}, {van Leeuwen}, \& {Zhu}}]{2016Natur.531..202S}
{Spitler}, L.~G., {Scholz}, P., {Hessels}, J.~W.~T., {et~al.} 2016, Nat, 531, 202

\bibitem[{{Tauris} {et~al.}(2013){Tauris}, {Sanyal}, {Yoon}, \& {Langer}}]{2013A&A...558A..39T}
{Tauris}, T.~M., {Sanyal}, D., {Yoon}, S.-C., \& {Langer}, N. 2013, \aap, 558, A39

\bibitem[{{Tavani} {et~al.}(2021){Tavani}, {Casentini}, {Ursi}, {Verrecchia}, {Addis}, {Antonelli}, {Argan}, {Barbiellini}, {Baroncelli}, {Bernardi}, {Bianchi}, {Bulgarelli}, {Caraveo}, {Cardillo}, {Cattaneo}, {Chen}, {Costa}, {Del Monte}, {Di Cocco}, {Di Persio}, {Donnarumma}, {Evangelista}, {Feroci}, {Ferrari}, {Fioretti}, {Fuschino}, {Galli}, {Gianotti}, {Giuliani}, {Labanti}, {Lazzarotto}, {Lipari}, {Longo}, {Lucarelli}, {Magro}, {Marisaldi}, {Mereghetti}, {Morelli}, {Morselli}, {Naldi}, {Pacciani}, {Parmiggiani}, {Paoletti}, {Pellizzoni}, {Perri}, {Perotti}, {Piano}, {Picozza}, {Pilia}, {Pittori}, {Puccetti}, {Pupillo}, {Rapisarda}, {Rappoldi}, {Rubini}, {Setti}, {Soffitta}, {Trifoglio}, {Trois}, {Vercellone}, {Vittorini}, {Giommi}, \& {D'Amico}}]{2021NatAs...5..401T}
{Tavani}, M., {Casentini}, C., {Ursi}, A., {et~al.} 2021, NatAs, 5, 401

\bibitem[{{Tendulkar} {et~al.}(2017){Tendulkar}, {Bassa}, {Cordes}, {Bower}, {Law}, {Chatterjee}, {Adams}, {Bogdanov}, {Burke-Spolaor}, {Butler}, {Demorest}, {Hessels}, {Kaspi}, {Lazio}, {Maddox}, {Marcote}, {McLaughlin}, {Paragi}, {Ransom}, {Scholz}, {Seymour}, {Spitler}, {van Langevelde}, \& {Wharton}}]{2017ApJ...834L...7T}
{Tendulkar}, S.~P., {Bassa}, C.~G., {Cordes}, J.~M., {et~al.} 2017, \apjl, 834, L7

\bibitem[{{Vanderlinde} {et~al.}(2019){Vanderlinde}, {Liu}, {Gaensler}, {Bond}, {Hinshaw}, {Ng}, {Chiang}, {Stairs}, {Brown}, {Sievers}, \& et~al.}]{2019clrp.2020...28V}
{Vanderlinde}, K., {Liu}, A., {Gaensler}, B., {et~al.} 2019, in Canadian Long Range Plan for Astronomy and Astrophysics White Papers, Vol. 2020, 28

\bibitem[{{Wang} {et~al.}(2022){Wang}, {Zhang}, {Dai}, \& {Cheng}}]{2022NatCo..13.4382W}
{Wang}, F.~Y., {Zhang}, G.~Q., {Dai}, Z.~G., \& {Cheng}, K.~S. 2022, NatCo, 13, 4382

\bibitem[{{Wang} {et~al.}(2023){Wang}, {Xu}, {Ma}, {Liu}, {Liu}, {Zhang}, {Pei}, {Chen}, {Manchester}, {Lee}, \& et~al.}]{2023SCPMA..6689512W}
{Wang}, N., {Xu}, Q., {Ma}, J., {et~al.} 2023, SCPMA, 66, 289512

\bibitem[{{Willmer}(1997)}]{1997AJ....114..898W}
{Willmer}, C.~N.~A. 1997, \apj, 114, 898

\bibitem[{{Xu} {et~al.}(2022){Xu}, {Niu}, {Chen}, {Lee}, {Zhu}, {Dong}, {Zhang}, {Jiang}, {Wang}, {Xu}, {Zhang}, {Fu}, {Filippenko}, {Peng}, {Zhou}, {Zhang}, {Wang}, {Feng}, {Li}, {Brink}, {Li}, {Lu}, {Yang}, {Caballero}, {Cai}, {Chen}, {Dai}, {Djorgovski}, {Esamdin}, {Gan}, {Guhathakurta}, {Han}, {Hao}, {Huang}, {Jiang}, {Li}, {Li}, {Li}, {Li}, {Li}, {Liu}, {Luo}, {Men}, {Niu}, {Peng}, {Qian}, {Song}, {Stern}, {Stockton}, {Sun}, {Wang}, {Wang}, {Wang}, {Wang}, {Wu}, {Xiao}, {Xiong}, {Xu}, {Xu}, {Yang}, {Yang}, {Yao}, {Yi}, {Yue}, {Yu}, {Yu}, {Yuan}, {Zhang}, {Zhang}, {Zhang}, {Zhao}, {Zheng}, {Zhu}, \& {Zou}}]{2022Natur.609..685X}
{Xu}, H., {Niu}, J.~R., {Chen}, P., {et~al.} 2022, Nat, 609, 685

\bibitem[{{Yamasaki} \& {Totani}(2020)}]{2020ApJ...888..105Y}
{Yamasaki}, S. \& {Totani}, T. 2020, \apj, 888, 105

\bibitem[{{Yao} {et~al.}(2017){Yao}, {Manchester}, \& {Wang}}]{2017ApJ...835...29Y}
{Yao}, J.~M., {Manchester}, R.~N., \& {Wang}, N. 2017, \apj, 835, 29

\bibitem[{{Zhang}(2023)}]{2023RvMP...95c5005Z}
{Zhang}, B. 2023, RvMP, 95, 035005

\bibitem[{{Zhang} {et~al.}(2020){Zhang}, {Yu}, {He}, \& {Wang}}]{2020ApJ...900..170Z}
{Zhang}, G.~Q., {Yu}, H., {He}, J.~H., \& {Wang}, F.~Y. 2020, \apj, 900, 170

\bibitem[{{Zhang} {et~al.}(2024){Zhang}, {Li}, {Zou}, {Zhao}, {Zhang}, \& {Zhang}}]{2024Univ...10..207Z}
{Zhang}, J.-G., {Li}, Y., {Zou}, J.-M., {et~al.} 2024, Universe, 10, 207

\bibitem[{{Zhang} \& {Zhang}(2022)}]{2022ApJ...924L..14Z}
{Zhang}, R.~C. \& {Zhang}, B. 2022, \apjl, 924, L14

\bibitem[{{Zhang} {et~al.}(2021){Zhang}, {Yan}, {Li}, {Zhang}, \& {Wang}}]{2021ApJ...906...49Z}
{Zhang}, Z.~J., {Yan}, K., {Li}, C.~M., {Zhang}, G.~Q., \& {Wang}, F.~Y. 2021, \apj, 906, 49

\bibitem[{{Zhao} {et~al.}(2021){Zhao}, {Zhang}, {Wang}, {Tu}, \& {Wang}}]{2021ApJ...907..111Z}
{Zhao}, Z.~Y., {Zhang}, G.~Q., {Wang}, Y.~Y., {Tu}, Z.-L., \& {Wang}, F.~Y. 2021, \apj, 907, 111

\bibitem[{{Zhu} {et~al.}(2023){Zhu}, {Niu}, {Cui}, {Li}, {Feng}, {Tsai}, {Wang}, {Zhang}, {Meng}, \& {Zheng}}]{2023Univ....9..251Z}
{Zhu}, Y., {Niu}, C., {Cui}, X., {et~al.} 2023, Universe, 9, 251

\end{thebibliography}

\end{document}